%% file: main.tex
\documentclass[preprint,journal]{vgtc}       % preprint (journal style)

\ifpdf%                                % if we use pdflatex
  \pdfoutput=1\relax                   % create PDFs from pdfLaTeX
  \usepackage{graphicx}                % allow us to embed graphics files
  \DeclareGraphicsExtensions{.pdf,.png,.jpg,.jpeg} % for pdflatex we expect .pdf, .png, or .jpg files
\else%                                 % else we use pure latex
  \usepackage{graphicx}                % allow us to embed graphics files
  \DeclareGraphicsExtensions{.eps}     % for pure latex we expect eps files
\fi%

\graphicspath{{figures/}{pictures/}{images/}{./}} % where to search for the images

\usepackage{microtype}                 % use micro-typography (slightly more compact, better to read)
\PassOptionsToPackage{warn}{textcomp}  % to address font issues with \textrightarrow
\usepackage{textcomp}                  % use better special symbols
\usepackage{mathptmx}                  % use matching math font
\usepackage{times}                     % we use Times as the main font
\usepackage{cite}                      % needed to automatically sort the references
\usepackage{tabu}                      % only used for the table example
\usepackage{booktabs}                  % only used for the table example
\usepackage{longtable}
\usepackage{enumitem}
\usepackage{hyperref}
\usepackage{tikz}
\usepackage{pgfplots, pgfplotstable}
\usepackage{flushend}
\usepackage{url}

\usepackage{savesym}
\usepackage{amsmath}
\savesymbol{amalg}
\usepackage{mathabx}
\restoresymbol{mabx}{amalg}

\usepackage{cleveref}

\preprinttext{To appear in IEEE Transactions on Visualization and Computer Graphics.}

\onlineid{1366}

\vgtccategory{Research}
\vgtcpapertype{application/design study}

\title{Designing Narrative-Focused Role-Playing Games \\ for Visualization Literacy in Young Children}

\author{Elaine Huynh, Angela Nyhout, Patricia Ganea, and Fanny Chevalier}
\authorfooter{
\item Elaine Huynh is with the Department of Computer Science at the University of Toronto. E-mail: elaine@cs.toronto.edu.
\item Angela Nyhout and Patricia Ganea are with the Ontario Institute for Studies in Education, University of Toronto. E-mail: \{patricia.ganea, angela.nyhout\}@utoronto.ca
\item Fanny Chevalier is with the Department of Computer Science and Statistical Sciences at the University of Toronto. E-mail: fanny@cs.toronto.edu.
}

\shortauthortitle{Huynh \MakeLowercase{\textit{et al.}}: Designing Narrative-Focused Role-Playing Games for Visualization Literacy in Young Children}
\usepackage{graphicx}
\usepackage{xcolor}

\newcommand{\visdel}[1]{}
\newcommand{\visadd}[1]{\textcolor{black}{#1}}

\newcommand{\cond}[1]{\texttt{\fontsize{8.5pt}{8pt}\selectfont #1}}
\newcommand{\TT}{C$\blacktriangleright$C}
\newcommand{\TF}{C$\blacktriangleright$I}
\newcommand{\FT}{I$\blacktriangleright$C}
\newcommand{\FF}{I$\blacktriangleright$I}

\abstract{Building on game design and education research, this paper introduces narrative-focused role-playing games as a way to promote visualization literacy in young children. Visualization literacy skills are vital in understanding the world around us and constructing meaningful visualizations, yet, how to better develop these skills at an early age remains largely overlooked and understudied. Only recently has the visualization community started to fill this gap, resulting in preliminary studies and development of educational tools for use in early education. We add to these efforts through the exploration of gamification to support learning, and identify an opportunity to apply role-playing game-based designs by leveraging the presence of narratives in data-related problems involving visualizations. We study the effects of including narrative elements on learning through a technology probe, grounded in a set of design considerations stemming from visualization, game design and education science. We create two versions of a game – one with narrative elements and one without – and evaluate our instances on 33 child participants between 11- to 13-years old using a between-subjects study design. \visadd{Despite participants requiring double the amount of time to complete their game due to additional narrative elements, the inclusion of such elements were found to improve engagement without sacrificing learning; our results indicate no significant differences in development of graph-reading skills, but significant differences in engagement and overall enjoyment of the game.} We report observations and qualitative feedback collected, and note areas for improvement and room for future work.}

\CCScatlist{ % not used in journal version
 \CCScat{K.6.1}{Management of Computing and Information Systems}%
{Project and People Management}{Life Cycle};
 \CCScat{K.7.m}{The Computing Profession}{Miscellaneous}{Ethics}
}

\teaser{
  \centering
  \includegraphics[width=\linewidth]{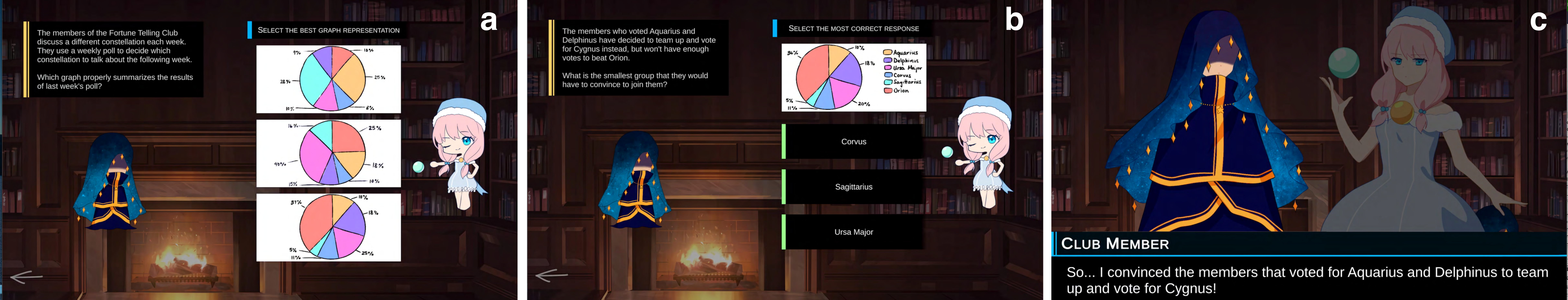}
  \vspace*{-2em}
   \caption{Sample screenshots from our educational role-playing game (RPG) to support visualization literacy education at ages 11 - 13. Our game encompasses data-related problems involving visualizations that young learners are required to solve to make progress in the game. \visdel{In this version, a multiple-choice question format is used.} Players must first indicate the right graph to solve the problem (a), and then interpret the selected graph to answer a question (b). \visadd{In this work, we investigate the effect of incorporating} narrative elements \visdel{are incorporated to foster}\visadd{on learning and} engagement (c). }
	\label{fig:teaser}
}

\renewcommand{\manuscriptnotetxt}{}

\vgtcinsertpkg

\begin{document}

\maketitle

\input{1-introduction.tex}
\input{2-related-work.tex}
\input{3-game.tex}

\input{4-methodology.tex}

\input{5-results}
\input{6-discussion-study}
\input{7-design-space}

\input{8-lessons-learned}

\input{9-limitations}

\input{9-conclusion}
\visadd{
 \acknowledgments{
We thank A. Iannuzziello and C. Reynier for input on tests and activities, W. Szeto for participating in analyzing textbook exercises, S. Hackeek for her help with recruitment, K.N. Truong, N. Riche, N. Sultanum and anonymous reviewers for their feedback on the manuscript. This work was supported in part by
 a grant from NSERC (RGPIN-2018-05072) and University of Toronto FA\&S Tri-Council Bridge Funding.}
}

\bibliographystyle{abbrv-doi}

\bibliography{biblio}
\end{document}

%% file: 1-introduction.tex
\section{Introduction}
From global-scale events such as climate change and public health crises, to societal topics such as economics, education, culture, or nutrition, many infographics, tables, and charts are created and circulated daily to help the general population better understand what is happening and what actions need to be taken. Graphical representations of data can exert considerable influence on the readership: the mere presence of a chart -- regardless of triviality -- is capable of convincing readers of the scientific credibility of an article \cite{Tal2016}. An important caveat is that, when designing data visualizations, parties may present information in ways \visdel{which}\visadd{that} mislead or deceive readers \cite{Pandey2015}, whether intentionally or not. As such, the ability to understand and appropriately handle data visualizations -- or more succinctly, \textit{visualization literacy} \cite{Boy:2014, Chevalier:2018} -- is a vital skill. Yet, despite its importance, visualization literacy amongst the general public remains low \cite{Borner:2016}.

% Paragraph 2: 
% The visualization community has recently started to recognize this gap 
% Prior work in elementary schools (Alper et al., Chevalier et al., Uta Hinrich)
% We add to these efforts, by exploring a novel form of educational support: gamification.

Recently, the visualization community has begun to acknowledge and address this deficiency. Children are exposed to charts and other data management topics as early as grade 1 \cite{OntarioCurriculum:2005}, and previous research has looked at ways to support visualization literacy education at the elementary school level \cite{Alper:2017, Chevalier:2018, Bishop:2020}. Research in this area is still in its infancy, and a range of methods to support learning, especially at a young age, have yet to be investigated. With our work, we add to these efforts through the exploration and application of a form of educational support not yet formally studied in visualization: gamification.

% Paragraph 3:
% Gamification has been used successfully in other topics.
% We see an opportunity to leverage this medium so as to support visualization learning at a young age
% There are many forms of games that can be designed - shortcomings of puzzles games etc....  [transition to our particular focus / motivation for that particular focus].

% As people become more connected \cite{StatsCan2017} and familiar with devices such as tablets and smartphones \cite{Papadakis:2013}, more effort is being put into finding ways of using technology to encourage learning even outside of the classroom. 

Gamification seeks to provide an interactive and engaging way of teaching players about new concepts or skills. We see gamification being applied to a broad range of subjects including music \cite{Chorlody} and physics \cite{LittleNewton} in an attempt to enrich the learning experience. Differences between the designs of educational games exist in order to better target difficulties and challenges offered by each topic. For example, gamification of basic arithmetic concepts such as addition or multiplication may result in a design \visdel{which}\visadd{that} encourages players to repeatedly solve problems in a short period of time. While this approach may work for some topics, this rote learning style of gamification has been criticized for its focus on training rather than actual learning \cite{Egenfeldt:2007, Gee:2003}, and may not lend itself well to tasks where memorization or recall are not key factors in understanding the material -- as is in the case with visualization literacy.

A review of publicly-available visualization problems posed to elementary school children (e.g., \cite{OntarioCurriculum:2005, IXL, KhanAcademy}) shows that stories are often used to provide context and motivation for questions related to chart interpretation. By leveraging the presence of such stories, we see an opportunity to apply a more narrative-focused approach to game design, such as those offered by role-playing games (``RPGs"). Our main research questions then become:
\vspace*{-0.5em}
\begin{itemize}%[noitemsep]
\setlength{\itemindent}{0.8em}
\setlength{\itemsep}{-0.2em}
    \item [\small{(RQ1)}]\visdel{How do we design}\visadd{What are the key factors to consider for} an RPG-style educational game focused on visualization literacy?
    \item [\small{(RQ2)}] How does the inclusion of narrative elements affect %children's
    learning and engagement, in the context of a visualization literacy game?
\end{itemize}
%\vspace*{-0.5em}
% Paragraph 4:
% In this work, we discuss the design consideration for the creation of such RPG-style educational game, grounded in the education literature + game design literature + narration. 
% We present an instance of the game, including exercises and narration.

% Paragraph 5:
% One key component of our work is whether the narration is important. 
% We design a study that evaluates two variations of the game: one with the narrative, and one without.

%\fanny{Rework this paragraph so it reflects the order of the paper.}
%In this work, we introduce the first story-based game to support visualization literacy learning, and discuss some considerations \visdel{which}\visadd{that} designers should be mindful of when creating an RPG-style educational game for visualization, grounded in the literature from both education and game design fields (\S\ref{sec:design}). We present and describe our own instance of a game, complete with a set of exercises, overarching narrative, and mechanics to support narrative elements (\S\ref{sec:game}), built as a technology probe to evaluate our approach. To gauge the impact of narrative elements in games focused on visualization literacy, we design a between-subjects study to evaluate two variations of our game: one with narrative elements, and one without (\S\ref{sec:study}).

In this work, we introduce the first story-based game to support visualization literacy learning. We \visdel{present and }describe our own instance of a game, complete with a set of exercises, overarching narrative, and mechanics to support narrative elements (\S\ref{sec:game}), built as a technology probe to evaluate our approach. \visadd{To gauge the impact of narrative elements in games focused on visualization literacy, we design a between-subjects study to evaluate two variations of our game: one with narrative elements, and one without (\S\ref{sec:study}). Our results (\S\ref{sec:results}) show that the inclusion of narrative elements yields a significantly positive impact on children's engagement and enjoyment, though it requires players to spend more time interacting with added elements. Interview sessions \visdel{with participants }reveal that children in the with-narrative condition are generally pleased with the story and associated interactions, but individual differences exist.}

\visadd{Using the qualitative feedback and observations from the study, we identify areas of improvement and discuss directions for future works (\S\S\ref{sec:discussion}-\ref{sec:conclusion}). Specifically, we contribute a set of design considerations to guide future work in creating RPG-style visualization games, grounded in the literature from both education and game design fields (\S\ref{sec:design-considerations}), and reflect on lessons learned (\S\ref{sec:lessons})}.

% Paragraph 6:
% Our results show that gamification is promising, and narrative is useful in our. We discuss directions for future works…

%% file: 2-related-work.tex
\section{Related Work}
Our work builds on prior literature in visualization literacy, educational games, and the role of narratives in games.

\subsection{Visualization Literacy}
%While graphs and visualizations have been around for a long time, it is only recently that the research community has defined the concept of \emph{visualization literacy}~\cite{Boy:2014, Chevalier:2018}

As studies surfaced the difficulties faced by the general population in interpreting and creating visual representations of data~\cite{Borner:2016,Kennedy:2014, maltese:2015}, ~\emph{visualization literacy} has emerged as an increasingly pressing concern. Visualization literacy, or visual information literacy~\cite{taylor:2003}, is the ability to critically read and construct data visualizations~\cite{Boy:2014, Chevalier:2018, Borner:2019}. Lacking such skill can be detrimental to individuals, resulting in poor decision-making through misinterpretation of charts. There is substantial anecdotal evidence of chart misuse, influencing the way people buy products, vote, or respond to a pandemic~\cite{cairo:2015, correll:2017}.

In response to this problem, visualization researchers have started to discuss a research agenda to promote visualization literacy, via a series of workshops \cite{Kim:2014, Romero:2014}, lessons learned from observational studies~\cite{Chevalier:2018}, or most recently, synthesis of accumulated knowledge in the domain~\cite{Borner:2019}. In an effort to develop\visdel{ed} standardized tools for measuring visualization literacy levels in people, Boy et al.~\cite{Boy:2014} introduced visualization literacy tests grounded in Item Response Theory, and Lee et al.~\cite{Lee:2006} proposed the Visualization Literacy Assessment Test (VLAT). Building the theoretical ground for guiding the design of a much needed visualization curricula has also been outlined as critical. Börner et al.~\cite{Borner:2019} contributed an initial proposal of data visualization literacy framework (DVL-FW), covering terminology and typology of core concepts in visualization, as well as a model to guide the development of curricula and assessment of visualization literacy. These research initiatives contribute to the goal of establishing structured, informed foundations, methods, models, and standardized instruments for assessing, teaching and learning visualization. 

One important concentration of effort in the area has been the documentation and research surrounding teaching and learning of visualization in young children. Alper et al.~\cite{Alper:2017} suggest that we can address visualization illiteracy in generations to come by improving the way we teach visualization as early as elementary school. To this end, they studied what materials are currently used in classrooms to support learning of visualization concepts through a review of textbooks and interviews with teachers. They designed ``C'est la vis", a technology-mediated intervention to support teachers as they teach pictographs and bar charts to students in grades K-4, aligned with current practices in education. Building on Huron et al.'s work on constructive visualizations~\cite{huron:2014a, huron:2014b}, Bishop et al.~\cite{Bishop:2020} studied the process of constructing visualization in children aged 5-12 using ``Construct-a-Vis", a tablet-based multi-user application where children can collaboratively or independently map visual variables to data dimensions in a free-form, creative manner. We continue these efforts focused on development of visualization literacy education in young children through the exploration of another form of intervention: gamification.

% \fanny{Conclude that we add to these efforts, by exploring a new form of interventions: gamification.}

\subsection{Gamification in Education}
While gamification long predates computer games \cite{Kafai2006}, the usage of educational video games (also referred to as ``serious games") is particularly appealing to younger generations. For such audiences, games have been found to be more engaging than passive media (e.g., videos) due to the presence of active control \cite{Greenfield:1984:MME:537939, Prensky}. Through games, players are able to engage in subject matter beyond simply reading and responding to problems, leading to increased performance when compared to traditional learning methods \cite{Papert:1991, Boeker:2013, Leutner:1993}. Games enable curiosity and experimentation through freedom to explore, allowing users to learn through experiential play \cite{Wang:2009}. In addition to the incorporation of active control, digital games take advantage of the auditory and visual streams enabled by electronic devices, and the combination of visualisation and exploration has been shown to enhance learning \cite{Betz:1995, Papert:1991, Amory:1999}. 

%Regardless of concerns expressed by a few over the efficacy of serious games\cite{Clark:1983, Egenfeldt:2007},  gamification remains a popular topic of study, spawning 
Interest in gamification has spawned a variety of educational games, covering broad range of topics such as physics \cite{LittleNewton} and music \cite{Chorlody}. In visualization\visdel{literacy}, there exists \visdel{games focused on bar charts \cite{Gabler:2019} and correlations in scatter plots \cite{GuessTheCorrelation} \visadd{(and also non-educational games where players evolve in an interactive infographics-looking landscape~\cite{metrico})}}\visadd{educational} games focused on bar charts \cite{Gabler:2019} and scatter plots \cite{GuessTheCorrelation}, \visadd{ as well as non-educational games inspired by infographics \cite{metrico}}. Yet, despite the variety of \visdel{game}genres and mechanics seen in commercial games, we observe that many of these educational games tend to focus on repetitive gameplay: players may be asked to complete a set of mathematical computations within a specified amount of time, or need to come up with a configuration to satisfy constraints for a particular level. \visadd{Some educational games integrate narrative elements to an extent (e.g., with short video clips or quests~\cite{Prodigy, brainpop, flocabulary}), but u}nlike the puzzle and action-based designs\visdel{adopted by many of these games}, there is a lack of research on the efficacy of \visadd{such} other genre and elements they offer\visdel{(e.g.,}\visadd{, especially }RPGs and their focus on storytelling, guidelines for designing such games, and information on how they affect learning and engagement.

\subsection{The Role of Narratives in Games and Education}
In commercial games, the presence of a story is credited with increased feeling of involvement in the gameplay and immersion in the environment \cite{Schneider:2004}. The presence of narratives in role-playing games helps to create memorable and pleasurable experiences, with players likening games in this genre to interactive story books and films \cite{Carr:2003}. Stories also act as intrinsic motivators, encouraging players to overcome obstacles and challenges in order to see what happens next in the story \cite{Gee:2003, Lieberman:2006, Wang:2009}. 

In online multi-player role-playing communities, there are often groups of players who engage specifically in role-playing activities \cite{Carr:2003}. These players create avatars in the game world, craft rich lore and personal stories for these characters, and interact with fellow role-players or roam the game space while assuming the identity of their creations. As such, it is clear that narratives are very important to at least a subset of players, and this should not be ignored when designing educational games.

In traditional education settings, the usage of story elements \visdel{have}\visadd{has} been shown to have positive effects on learner memory, motivation, and engagement \cite{Bielenberg:1997}, though the degree of these effects may vary between groups based on level of prior knowledge \cite{Wolfe:2007}. In teaching higher-level math and computer science concepts, incorporating narratives and storytelling has been suggested in order to generate interest in the subject matter \cite{Papadimitriou}. There has also been an attempt at interleaving storytelling and learning in the context of a textbook \cite{Field:2016}, which has been generally well-received by the public and praised for its engaging presentation of educational content.

When looking at the effects of narratives in serious games specifically, there seems to be conflicting information with respect to how conducive they are to learning. Adams et al. experimented with the effects of narratives in two games and concluded that, compared to simply viewing a slideshow presentation, playing a narrative game does not aid in learning \cite{Adams:2012}. In contrast, Bittick reported that narratives increased self-perceived immersion in serious games and that gender plays a role in the effectiveness of narratives on learning outcomes \cite{Bittick:2011}. Another study \cite{Fokides:2019} found that narration quality had positive effects on learning effectiveness, as well as enjoyment. Further work looking at narratives in serious games is required before making claims as to whether or not they are largely beneficial or detrimental to learning.

%% file: 3-game.tex
\begin{figure*}[t!]
 \centering
 \includegraphics[width=\columnwidth]{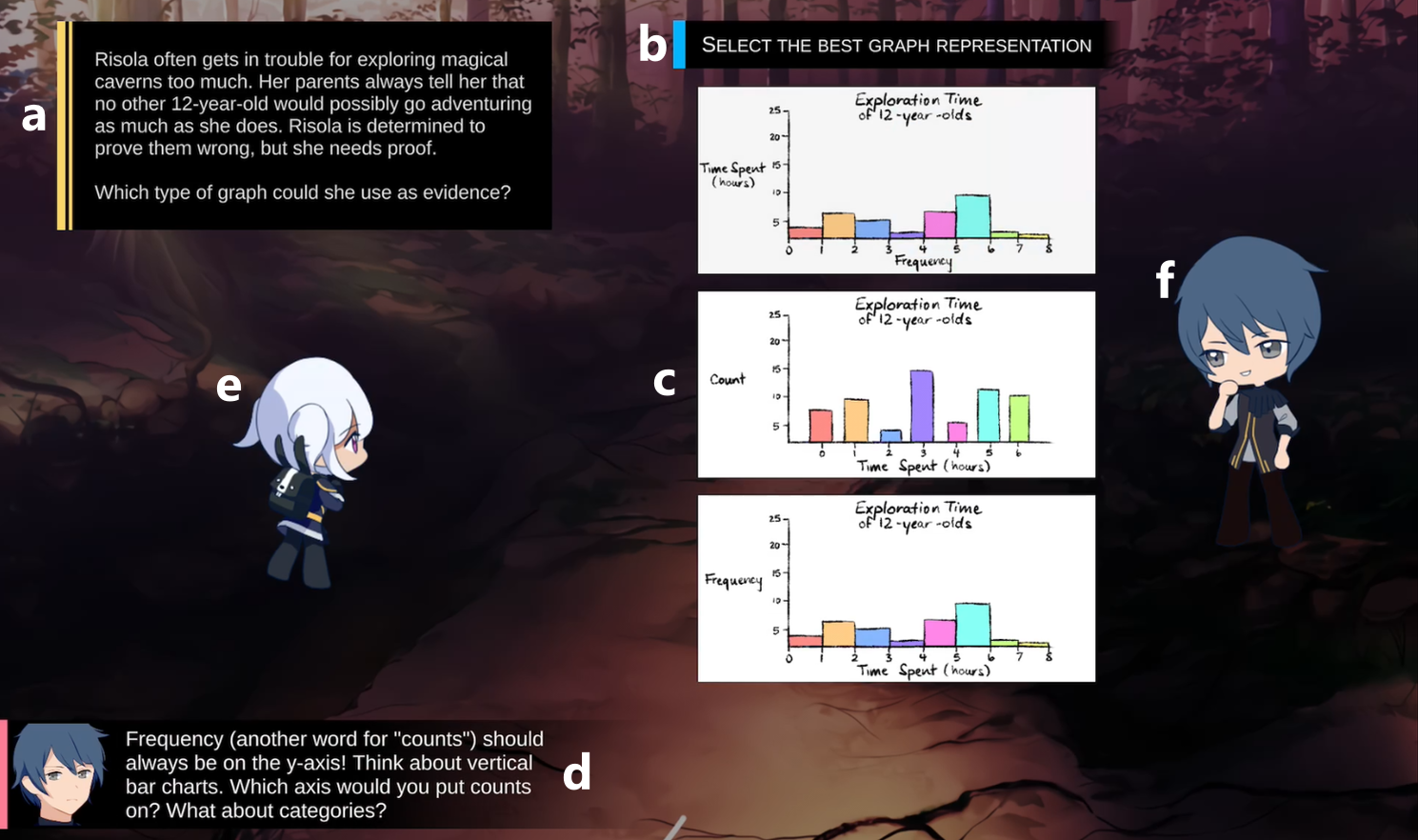}
 \includegraphics[width=\columnwidth]{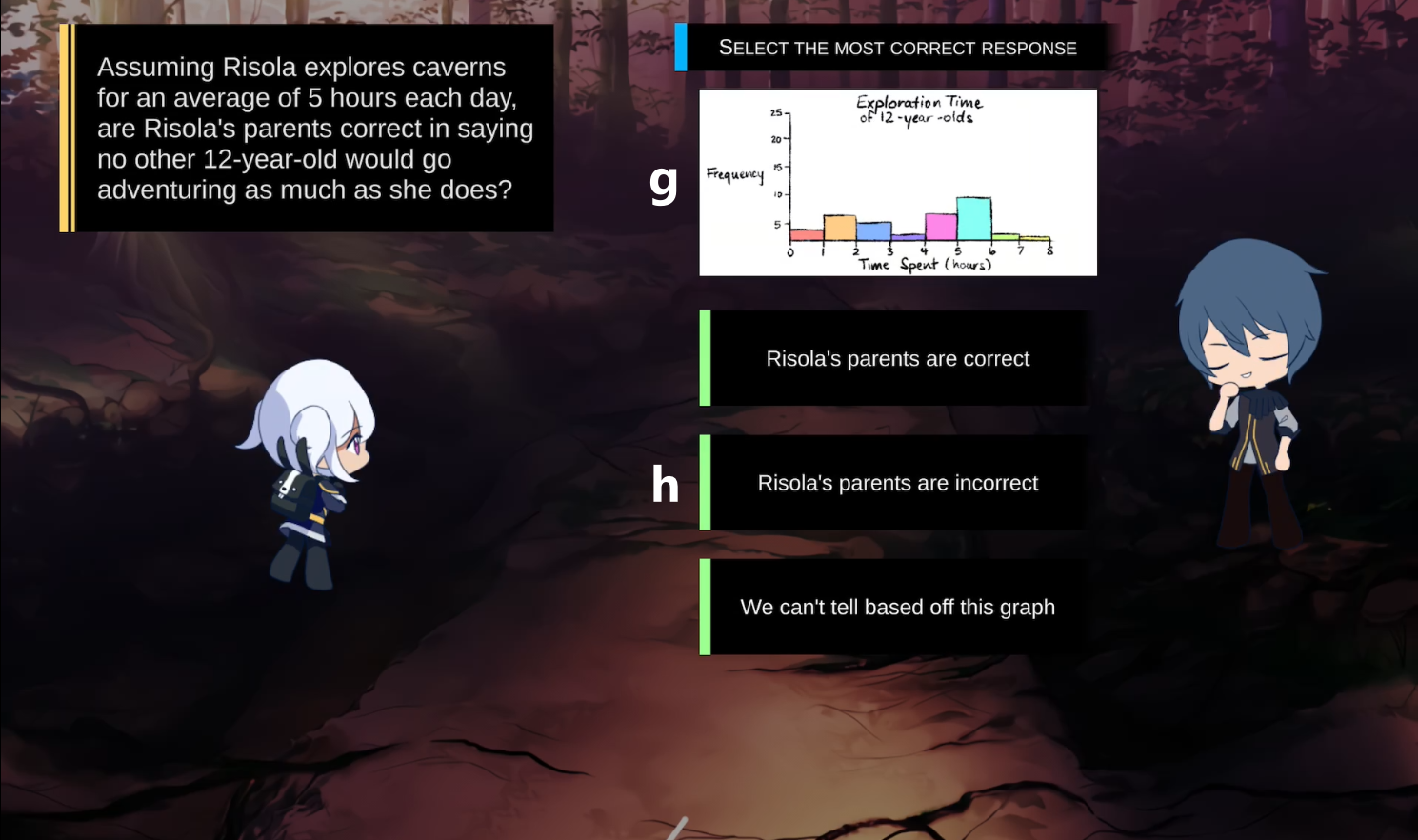}
  \vspace*{-1em}
 \caption{\textit{Activity View}. This view consists of a question display box (a), instructions (b), choices to select from (c), a feedback box (d), and character sprites (e, f). Follow-up ``interpret the chart" questions include a picture of the previously-chosen chart (g), and choices of written answers (h).}
 \label{fig:activity-view}
  \vspace*{-1.5em}
\end{figure*}

\section{Technology Probe: A Game to Learn Visualization}
\label{sec:game}

%\nath{I dont like proof of concept. perhaps Technology Probe: A Game to learn Vis}
% \elaine{"visualizations"? but it's too long for one line...}
% \fanny{visualization, as the discipline/area}
%  ok

We first briefly describe our own example of an educational RPG, the design decisions we made, and its implementation. This technology probe was created with the intent of studying the impact of narrative elements in a visualization-focused educational game on learning and engagement \visadd{(RQ2; see \S\ref{sec:study})}. We share further reflections on design considerations to account for when designing such games in \S\ref{sec:design-considerations}.

\subsection{Target Audience}
For our game and subsequent study, we targeted children between 11- to 13-years old, as we felt they would be comfortable with the amount of reading required in an educational RPG. \visdel{; targeting a younger age may impact findings due to differences in reading ability. In addition, our selected}\visadd{This} age range \visadd{also} includes the intersection of children who are most likely to read for fun in their spare time, as well as children who already express familiarity with electronic games and applications \cite{Scholastic:2017}. 

\subsection{Educational Activity Design}
\label{sec:activities}
%To guide the design of our educational activities, 
We first looked at the provincial curriculum \cite{OntarioCurriculum:2005} to understand which topics our participants should be familiar with and decided our educational activities would focus %on two topics: 
on pie charts (i.e. ``circle graphs"~\cite{OntarioCurriculum:2005}) and histograms. %We chose circle graphs because it seemed likely that our participants would be somewhat familiar with them, but might still find the topic challenging. Histograms, on the other hand, were likely to be new to the majority of them. This would allow us to study the game as a tool for practice or rehearsal of previously-taught topics, as well as a supplementary tool for teaching new concepts.
Pie charts were chosen because children in this age range should be mostly familiar with them but still find them challenging, testing our game as a supplementary tool for practice %and rehearsal 
of previously-taught topics. Histograms would likely be new to most participants, giving us insight on the game as a tool for teaching new concepts.

Educational activities are presented as multiple-choice questions, consisting of \textbf{choose a chart} question (i.e. which chart is best suited and/or the correct one to solve the problem), followed by an \textbf{interpret the chart} question. To further streamline the process through which players would be exposed to questions, we presented these question types in pairs (e.g. Figure~\ref{fig:teaser}a, then Figure~\ref{fig:teaser}b).

%Regarding question types, 
%\nath{I would move all the decision making into next section - so this paragraph moves. you make it more succinct here and enable reproduction. and you make the next section more interesting.}
%We opted for multiple-choice questions due to the ease of implementation, verification, and their ability to support a breadth of topics, and looked at sample exercises and worksheets \cite{TIPS4RM, IXL, KhanAcademy} for guidance. We noticed many chart questions could be classified further into either create a chart-type questions, or interpret the chart-type questions. The latter would be simple to translate into multiple choice questions, %(i.e., by providing possible interpretations as options), 
%but the former would present a challenge unless players are asked to make a choice for each part of the chart (e.g., ``Which of the following options is best suited as a label for the y-axis in this chart?"). To address this, we made use of questions where players would be presented with different charts and have to choose the one which best fits a scenario% or would be the most useful in helping someone answer a question
%. 

We designed 10 story-neutral question templates (5 pairs) for each of the charts tested, resulting in a total pool of 20 questions. These templates captured the learning objective/topic to be tested (e.g., understanding intervals in histograms) in order to mimic the structure of questions found in typical textbook exercises, and omitted the presence of a story or scenario, which would be added after the narrative design was complete. For each question template, we created three possible options/responses, along with textual feedback to be displayed whenever an option is selected as a response. The feedback provided was intended to provide hints and educate players on why their answer may be incorrect, as well as provide further explanation or reinforce understanding for cases where players chose the correct answer. Two elementary school educators teaching in Canada and the US, respectively, helped to review and refine the generated exercises and feedback. 

% \fanny{Do we want to mention William here too? Did he help "validate" the content?} \elaine{He mainly helped with the initial categorization of questions in 2.2.}
% \fanny{Do you mean in the following that we first created a story-neutral "template" of the problems? This should be more clearly conveyed. Also, the argument is \emph{not} that the integration would be easier. Rather, it was for us to guarantee that our exercises complied with typical exercises as found in textbooks.} 

\subsection{Narrative Design}
\visadd{To add narrative context to the activities, players are further immersed in a game world that they can \textbf{explore} and find characters to interact with through \textbf{dialogues}}. We opted for a linear narrative to allow for stronger control over how players interact with the story in our study \visadd{(see \S\ref{sec:narrative-design})}. A simple story with a dramatic arc was used and situated in the fantasy genre to mimic those seen in generic RPGs. Players are first introduced to the main character -- a student at a magic school -- and assume her role as she meets other characters, helping them solve their problems. We used the narrative to set up context for each of the educational activities, and reiterated the problem in the question text to be displayed when solving a problem to promote fairness between conditions. In linking the story and educational content, we loosely tied the difficulty of problems with parts of the dramatic arc: histogram questions, which we believed would be more difficult, were presented only in the latter part of the game, during the rising action and climax of the dramatic arc. We hoped that, while players may be discouraged by the difficulty of the questions, they might be motivated and encouraged to persevere to see the rest of the story, as the plot becomes more interesting and intense.

%For example, a multiple-choice question related to determining the best spellbook to sell would be prefaced in the narrative with dialogue about a character who is interested in setting up a stall and is unsure about what to sell.

\subsection{Game Design}
\label{sec:game-design}
Our proof of concept was made using the Unity game engine and compiled for use on 10.5" iPad Pro. %due to the increased screen size and ease of observation over smartphones, and improved portability over PCs and laptops. 
%\nath{this paragraph goes to next section}
The game consists of three main views: the \textbf{Activity View} (Figure \ref{fig:activity-view}) \visadd{presenting the puzzles}, \visadd{and} the \textbf{Dialogue View} (Figure \ref{fig:teaser}c)\visdel{,} and \visdel{the}\textbf{Exploration View} (Figure \ref{fig:exploration-view}) \visadd{to support the narrative dimension of the game}. Through the Activity View, players are presented with the chart-related questions (\S\ref{sec:activities}). Upon selection of a response, a feedback box (Figure \ref{fig:activity-view}d) appears in the bottom-left corner of the screen with a predefined message associated with that choice. Players are permitted (without penalty) to make choices until they choose the correct answer, after which the Activity View ends and the next view is loaded. \visadd{The Activity View is typical of many educational games \cite{LittleNewton, Chorlody, Gabler:2019, GuessTheCorrelation}, where puzzles are presented in sequence without a narrative connecting them together}.

\visadd{In our RPG-like game, p}layers are also able to interact with the Dialogue and Exploration Views. Progression is relatively streamlined: players engage with the narrative through character dialogue \visadd{(via Dialogue View)} and \visdel{then }go through a phase of exploration \visdel{(via Exploration View) and character interactions (via Dialogue View)}until they interact with the character who will advance the story for them \visadd{and present the question in the Activity View.}\visdel{. In doing so, they are presented with a question in the Activity View.} Once finished, players are sent back to the Dialogue View to wrap up the conversation and be given the next objective. This repeats until the player fulfills all objectives. Across all views, interaction is done through simple tapping.

%: players indicate their answer to questions by tapping on choices, proceed through dialogue by tapping anywhere on the screen, move between maps by tapping on marked areas, and initiate character interactions by tapping on a character's sprite.

%% file: 4-methodology.tex
\section{Comparative Study: The Role of Narrative}
\label{sec:study}

%\begin{figure*}
% \centering
% \includegraphics[width=\columnwidth]{pictures/study-setup.jpg}
% \includegraphics[width=\columnwidth]{pictures/child-interaction.png}
% \caption{Setup of the room used for testing (left) and a child interacting with our game (right).}
% \label{fig:setup}
%\end{figure*}

Using a between-subjects study design, we evaluated \visadd{(RQ2):} the impact of \visadd{incorporating} narrative elements \visadd{(i.e., exploration and dialogues)} in educational RPGs on learning and engagement on children aged 11-13, using our design probe (\S\ref{sec:game}). Participants were tested in independent % of one another, in a session c
sessions conducted in a \visdel{private }lab setting with only the participant and investigator present, for a maximum of one hour. \visdel{Before testing commenced, p}\visadd{P}articipants were informed that they could stop at any time without penalty, and were permitted to ask questions %or otherwise interact with the investigator 
if they needed any assistance. % or clarification. %In each testing session,
\visadd{The study consisted of four phases: Pre-Test, Play Time, Post-Test, and Interview. In Play Time, p}articipants \visadd{were all presented with the same activities and} were randomly assigned to either the \cond{without-narrative} (i.e., \visdel{educational}activities only\visadd{, presented in sequence without further context}), or \cond{with-narrative} (i.e., \visdel{all content, including }\visadd{activities,} exploration, and dialogues\visdel{ included}) condition\visdel{, and led through four phases: Pre-Test, Play Time, Post-Test, and Interview}. Children were compensated \visdel{for their time }with their choice of small toys; parents received monetary compensation for travel expenses.

%Pre-Test (\S\ref{sec:pretest}), Play Time (\S\ref{sec:Play Time}), Post-Test (\S\ref{sec:posttest}), and Interview (\S\ref{sec:interview}). Children were compensated for their time with their choice of small toys; parents received monetary compensation for travel expenses.

\paragraph{Participants.}
Participants were recruited through a database of families who had previously expressed an interest in child studies and collected over several years using mailing lists, public events, and word of mouth. Parents of children between 11 to 13 were contacted by phone or by e-mail, and given a short description of the study\visdel{along with information regarding compensation}. Due to the amount of reading involved in the study, we requested that only children who could read at a grade 6 level or higher participate in the study. A total of 33 participants (16 female, \visadd{mean} age of 12.09) participated in the study. All recruited participants were native English speakers, with two having attended a French immersion school.

%\begin{figure}[htb]
% \centering
% \includegraphics[width=\columnwidth]{pictu%res/narrative-3-cropped}
% \vspace*{-2em}
% \caption{\textit{Dialogue View}. The name of the active speaker and their text are rendered in the dialogue box at the bottom of the screen while portraits indicate which characters are involved in the conversation. Focus is drawn to the active speaker by darkening inactive ones.}
% \label{fig:dialogue-view}
%  \vspace*{-1em}
%\end{figure}

\paragraph{Pre-Test.}
\label{sec:pretest}
Following introductions, the child participant was given a test to gauge their prior understanding of pie charts and histograms. \visdel{We consulted with a professional educator about the design of the pre-test and made adjustments to address concerns such as the amount of time it would take for participants to complete each question and ease of grading. The pre-test was also modified to include basic arithmetic and reading comprehension questions to gauge competence.} \visadd{We co-designed the pre- and post-tests with a professional educator, and iteratively refined the instrument to take into account the amount of time it would take for participants to complete each question, ease of grading, as well as learning objectives. In particular, the educator suggested to include basic arithmetic and reading comprehension questions to gauge competence.} Questions using both fantasy and non-fantasy contexts were included to address transfer of knowledge between domains. The final version of the pre-test distributed to participants consisted of one page of questions gauging basic math and reading comprehension skills (4 addition, 4 subtraction, 3 multiplication, and 2 simple word problems), followed by 6 pages of chart problems (8 multiple-choice questions). Table \ref{tab:question-chart} outlines the order and types of chart questions used.

\paragraph{Play Time.}
\label{sec:Play Time}
Participants were asked to play through one version of the game (\cond{with-narrative} or \cond{without-narrative}). This phase was video- and screen-recorded for further analysis. %, and participants were made aware of this prior to recording. 
Children were told they would have approximately 30 minutes to complete the game, but there would be no penalty or reward associated with speed of completion. We quietly observed participants, %as they engaged with their assigned game, 
and made notes about whether they appeared to think about the questions or brute-forced solutions, if they read the feedback box, and how they interacted with the story (e.g., if they appeared to read through all dialogue or skipped text). The investigator only intervened when the participant appeared to be stuck or if they requested clarification or assistance.

\begin{figure}[tb]
 \centering
 \includegraphics[width=\columnwidth]{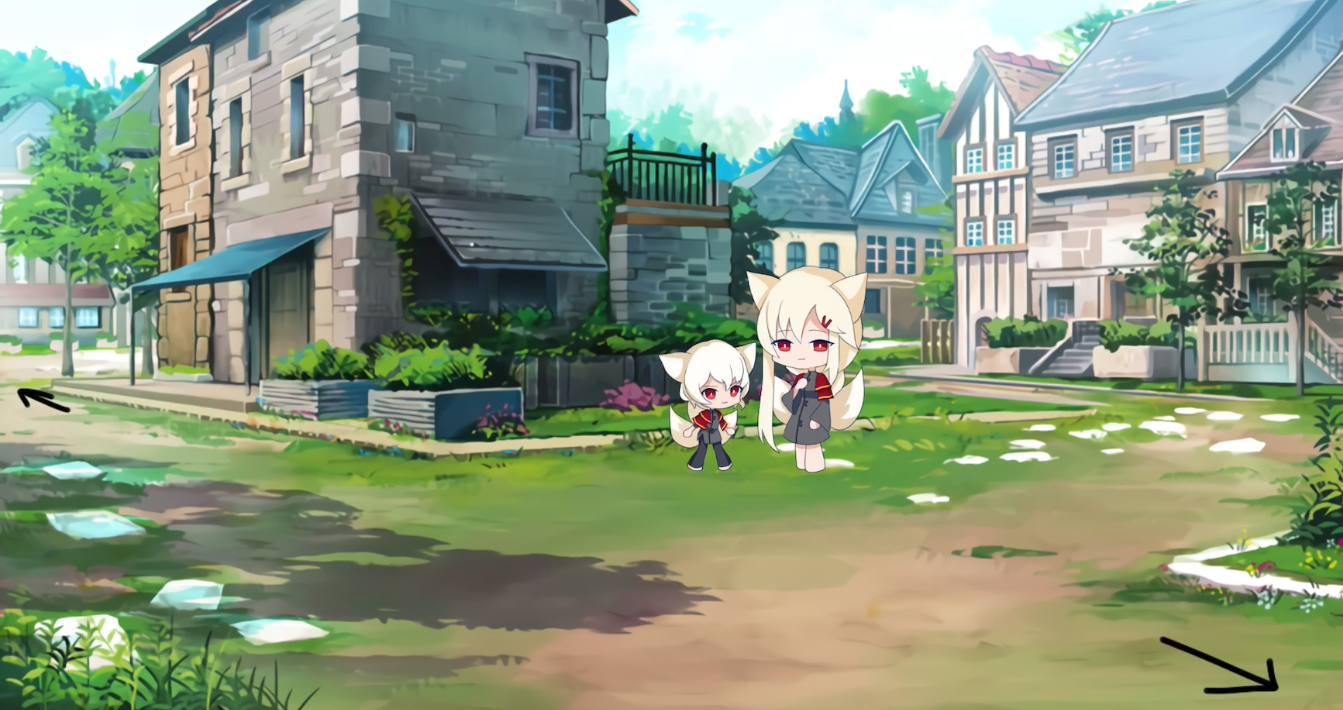}
  \vspace*{-2em}
 \caption{\textit{Exploration View}. Players are able to explore the game world by tapping on arrows (hotspots) connecting to adjacent maps. Character interactions are triggered by tapping on a character's respective sprite.}
 \label{fig:exploration-view}
  \vspace*{-1em}
\end{figure}

\begin{table}[tb]
 \centering
 \small
 \begin{tabular}{r  c  c  c}
     & \textbf{Chart} & \textbf{Question Type} & \textbf{Scenario} \\
    \hline
     Q1. & Pie & Choose a Chart & Non-Fantasy \\
     Q2. & Pie & Choose a Chart & Fantasy \\
     Q3. & Pie & Interpret the Chart & Non-Fantasy \\
     Q4. & Pie & Interpret the Chart & Fantasy \\
     Q5. & Histogram & Choose a Chart & Non-Fantasy \\
     Q6. & Histogram & Choose a Chart & Fantasy \\
     Q7. & Histogram & Interpret the Chart & Non-Fantasy \\
     Q8. & Histogram & Interpret the Chart & Fantasy \\
 \end{tabular}
 \vspace*{0.5em}
 \caption{Order and classification of chart questions used in the tests.}
 \label{tab:question-chart}
  \vspace*{-2em}
\end{table}

\paragraph{Post-Test.}
\label{sec:posttest}
Participants were then given a test on paper to gauge changes in their understanding of the subject matter. This post-test was identical to the one provided in the pre-test, but with the basic math and reading comprehension questions removed.

\paragraph{Post-Study Interview.}
\label{sec:interview}
\visdel{At the end of the testing session, w}\visadd{W}e conducted semi-structured interviews\visdel{with participants} to gather feedback on \visdel{the design of our game}\visadd{our approach}. Participants were \visdel{first }asked to rate \visadd{and share their general thoughts about} different aspects of the game such as the art, educational activities, engagement, fun, and overall gameplay. Plot and characters were also included for those \visdel{who were assigned to }\visadd{in} the \cond{with-narrative} condition. \visdel{We also gathered qualitative feedback.
Participants in the \cond{with-narrative} condition were first asked about their general thoughts on the story and characters.} In cases where \visdel{they}\visadd{participants} were unresponsive, the investigator attempted to obtain more information by asking the participant to \visdel{separately }list what they enjoyed and did not enjoy about the topic. Children in this condition were then asked whether they noticed themselves skipping through the story or dialogue, and, if so, explain why they might have engaged in such behaviour. \visdel{Beyond this point, all questions asked were the same between both conditions.}

We \visadd{then} asked participants \visadd{from both groups about}\visdel{ for} their feelings on the educational activities, if they engaged in brute-forcing strategies, and why they might have done so\visdel{. While we did record observations regarding whether each participant skipped the story or brute-forced answers, we asked p}\visdel{Participants \visadd{were asked} to self-report if they engaged in such activities as opposed to directly asking them to explain themselves, and only \visdel{asked}\visadd{pressed} for additional information} when they were comfortable \visdel{enough }with acknowledging this behaviour. Finally, thoughts \visdel{about}\visadd{on the usefulness of} the feedback box\visadd{,}\visdel{ and whether or not they found it to be helpful were solicited, along with thoughts on} the educational value of the game, \visadd{and} suggestions \visadd{on}\visdel{for} how to improve \visadd{ were solicited}. \visdel{, and any additional feedback or comments they may have.}

%% file: 5-results.tex
\section{Results}
\label{sec:results}

\begin{figure}[t!]
    \centering
    \includegraphics[width=0.7\linewidth]{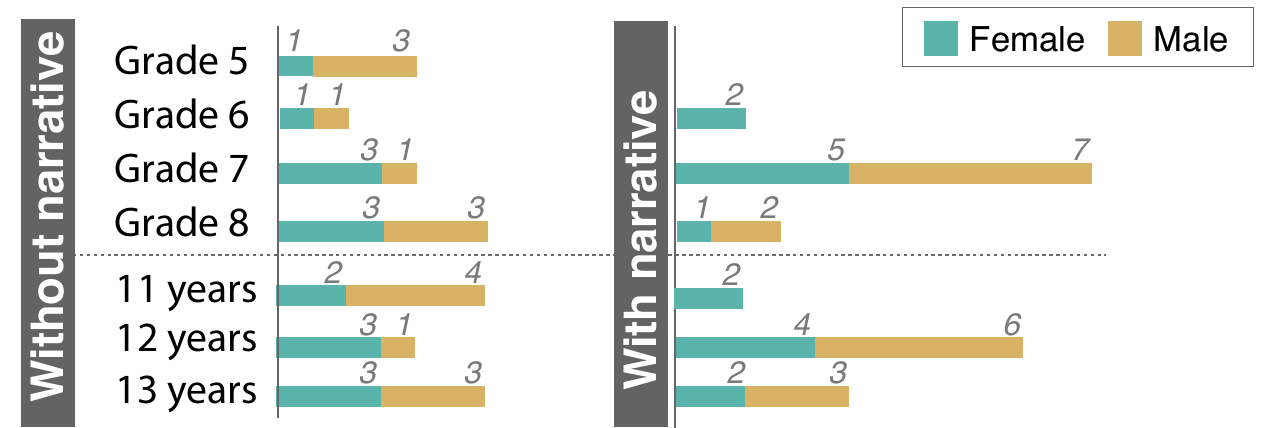}
    \vspace*{-1.2em}
    \caption{Demographics of participants per condition.}
    \vspace*{-1em}
    \label{fig:demographics}
\end{figure}

Child participants were assigned to the \cond{without-narrative} condition (16 participants\visdel{, 8 female\visadd{, 8 male}}) or the \cond{with-narrative} condition (17\visdel{ \visadd{participants}, 8 female\visadd{, 9 male}})\visadd{; see demographics breakdown in Figure~\ref{fig:demographics}}. Analysis of the basic comprehension test showed that, at most, participants made a single mistake (9/33, 5 \cond{with-narrative}, 4 \cond{without-narrative}). From this, we determined all participants were fit to grasp the study material and results of the study were unlikely to be affected by fundamental differences in cognitive ability. All participants played their assigned games to completion, spending an average of 13min (\cond{without-narrative}) and \visdel{25min (\cond{with-narrative}) effectively playing the game. Participants in the \cond{with-narrative} condition spent an average 10min on educational activities, 13min on dialogue, and 2min on exploration.}\visadd{10min (\cond{with-narrative}) on educational activities. Participants in the \cond{with-narrative} condition spent an additional 13min on dialogue and 2min on exploration, for a total of 25min.} Interviews lasted about 7-8min. One participant (P25) did not complete the interview due to time constraints. Video and audio recordings were transcribed for analysis\visdel{, and we now discuss main findings relative to our research question \visadd{RQ2}}.

\begin{figure}
    \centering
    \includegraphics[width=\linewidth]{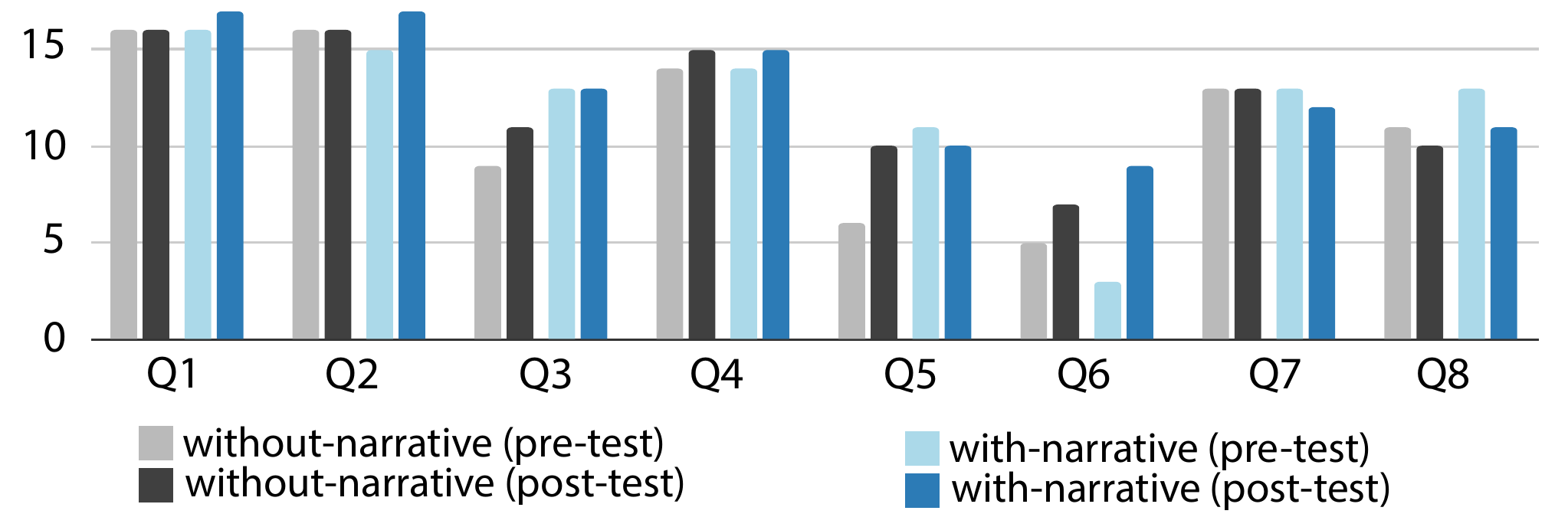}
    \vspace*{-2.2em}
    \caption{Total number of participants answering questions correctly in the pre- and post-tests and across conditions}
    \vspace*{-1.2em}
    \label{fig:breakdown}
\end{figure}

\subsection{Impact on Learning}
\paragraph{Quantitative Analysis.} Figure~\ref{fig:breakdown} shows the number of participants responding correctly to questions in the pre- and post-test across conditions. For each condition, we conducted a paired-samples t-test to compare the performance in the pre-test and post-test for each question. There was a significant difference in the pre-test scores (M=0.267, SD=0.449) and post-test scores (M=0.485, SD=0.507) for \visdel{only }one question (Table~\ref{tab:questions}, Q6), in the \cond{with-narrative} condition; t(16)=0.436, p = 0.0093. We did not find any other significant differences\visdel{. That said}, \visadd{but} for some questions (e.g.\visadd{,} Q1), there was no room for \visdel{possible }improvement as most, if not all, participants answered the question correctly in the pre-test (Fig.~\ref{fig:breakdown}). 

For each pair of pre- and post-test questions, there were four possible outcomes a participant could achieve, whether they would answer the question correctly (C: correct) or not (I: incorrect) on the pre- then post-test: \TT, \TF, \FT, \FF. Focusing on the overall changes in performance at a per-participant level, we now discuss results based on the number of times participants could have improved (sum of \FT~and \FF), showed improvement (total \FT), or did worse (total \TF).

\noindent $\bullet$ \textbf{Gender.} Grouping by gender revealed that all 8 males in the \cond{without-narrative} condition either improved (\FT $>$ 0; 6/8 males) or showed no change between tests (\FT = 0; 2/8). Cases where males performed worse (\TF $>$ 0; 4/8) were only found in the \cond{with-narrative} condition. There were no such trends seen in female participants.

\noindent $\bullet$ \textbf{Age \visadd{and Grade}.} All 11-year-old participants \visdel{(8/33\visadd{; 4 in grade 5, 4 in grade 6}) }either improved (6/8\visadd{; 3 in grade 5, 3 in grade 6}) or showed no change (2/8\visadd{; 1 in grade 5, 1 in grade 6}) between tests. None performed worse on any of the questions. \visdel{There were n}\visadd{N}o clear patterns \visadd{were apparent} in \visdel{any }other age\visadd{/grade} groups.

\noindent $\bullet$ \textbf{Degree of Change.} Participants who exhibited the greatest change between tests (3/33) were observed to belong to the \cond{with-narrative} condition (\FT = 3, 2/3; \TF = 1, 1/3). 

%\nath{why did you test for these groups? kind of missing the hypotheses I think. Did you expect these?  also for degree of change, that is not much 3 people?}

\paragraph{Qualitative Analysis.} When asked about the educational value of the games they played, most participants felt the game would be helpful to people who struggle with charts or understanding word problems. %\nath{so they find it helpful even though they did not improve their results? that is interesting...} 
Some specifically noted that the activities would have been better suited for students who are new to the covered charts, as opposed to someone who is already familiar with and wanting to learn more about them. This finding aligns with the quantitative data above regarding performance based on age, as children in the 11-year-old group were less likely to be familiar with concepts compared to older participants.

\begin{figure}[t!]
    \centering
    \includegraphics[width=\linewidth]{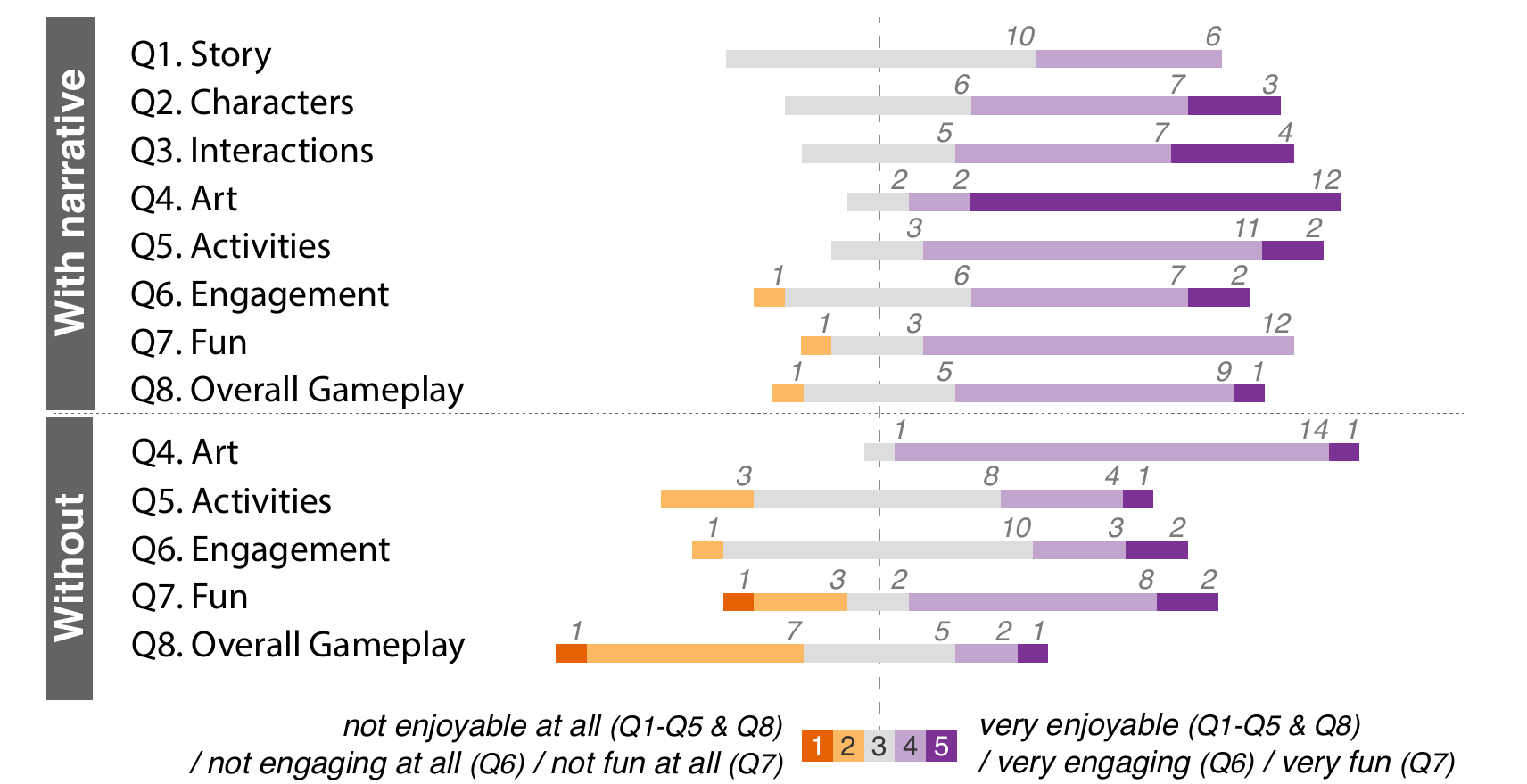}
    \vspace*{-2em}
    \caption{Participant ratings from the interview phase.}
    \label{fig:ratings}
    \vspace*{-1.3em}
\end{figure}

Regarding the feedback box --- whose primary goal was to help learners think critically when making mistakes --- the majority of participants (21/32) considered the functionality to be helpful or conducive to learning, likening it to \textit{``having a teacher beside you."} (P1) as \textit{``it gave you little recaps about how your answer was and how it thought you did"} (P15), which was found particularly valuable \textit{``when the question was hard"} (P15). Still, a few participants (P4, P7, P12, P23) felt conflicted about the overall usefulness of the feedback because they found it to be not as helpful when they made a correct choice, commenting \textit{``it was like a pat on the back"} (P4). In contrast, P17 noted that the feedback \textit{``made [her] feel happy because whenever [she] got a question right, they said `good job'."} Another participant (P14) mentioned that they would \textit{``half-read"} the feedback box when their answer was correct.

Four participants did not find the feedback box to be helpful (P13) or simply did not bother to read it (P16, P22, P27), but could not articulate why they felt this way when asked to elaborate. Two participants personally considered the feedback box to not be an important feature, but mentioned they would prefer to keep it rather than remove it, if given the choice (P19), and \textit{``for some people who don't understand certain parts of math, it would help them"} (P21).

\subsection{Impact on Engagement}
\paragraph{Quantitative Analysis.} Figure~\ref{fig:ratings} shows the results of the subjective ratings collected during the interview. Participants in the \cond{with-narrative} condition appeared to be more engaged and content with the game than those in the \cond{without-narrative} condition. %\nath{but is engagement good or bad? could it be distraction? need to be discussed I think} 
We ran Mann-Whitney's U tests on each rated category and found significant differences between the two conditions for: art (U = 53, Z = -3.14, p = 0.002, r = 0.56), activities (U = 66.5, Z = -2.49, p = 0.01, r = 0.44), engagement (U = 63, Z = -2.72, p = 0.007, r = 0.48), and fun (U = 57.5, Z = -2.82, p = 0.005, r = 0.50).

\paragraph{Qualitative Analysis.} Thoughts on the story and character were mostly positive, with some participants praising the usage of different facial expressions on characters to express emotions, as opposed to displaying a single static image per character. Two participants (P22, P29) explicitly stated they enjoyed how the story and the problems were connected to each other, and one participant (P31) liked how the overarching narrative provided motivation behind the problems:

\begin{quote}
\vspace*{-.4em}
\fontsize{8.5pt}{8.5pt}\selectfont
    \textbf{P31:} \textit{``They're questions that are actually in-depth and it's not just like, 'This person has this, that person has that, my answer is this.' Why would I care how much stuff Joan has? Why do I care about this person's problem? ... This game is like, 'I need your help, can you tell me what this is?' and less like 'This person needs this thing, now explain what they need to do.'"}
\end{quote}

Unprompted, one participant (P10) in the \cond{without-narrative} condition stated, \textit{``The questions didn't really follow a story. I'd kinda like to fit it in a story."} However, not all feedback was positive. Some participants (P4, P5, P14) felt that the story and interactions were too short and would have benefited from more detail, commenting that \textit{``a lot of the characters were very shallow -- like not much personality."} (P4), and that it was regrettable that \textit{``for some of the characters, you talked to them once and never saw them again."} (P14)

Finally, while most participants enjoyed the fantasy aspect of the narrative, two participants (P7, P15) remarked that the usage of fantasy contexts -- whether they are interesting or not -- made the problems more difficult to them:
\begin{quote}
\vspace*{-.4em}
\fontsize{8.5pt}{8.5pt}\selectfont
    \textbf{P07:} \textit{``I got a bit confused about how they kept talking about spells because I'm just not engaged with that overall so I wasn't really interested in that, but I think the questions -- because they used spells -- would both be engaging for people who do enjoy it, and in a way, used in the real world."} %There was one part where there were caves and they were like, ``Oh, kids your age aren't using it", and that's, like, ``Oh, kids your age aren't using their phones often", so like... It kinda involves the real world as well."}
\vspace*{-.1em}  
\end{quote}

\begin{quote}
\vspace*{-.4em}
\fontsize{8.5pt}{8.5pt}\selectfont
    \textbf{P15:} \textit{``I felt the drawings [...] % -- like the art and the background -- 
    kinda engaged you and the questions were really mystical and interesting types of questions. I think it would've been slightly easier, in fact, with questions about the real world, but I think that's why most mathematical games like Prodigy\cite{Prodigy} veer off to mythical questions -- because they're a bit more difficult to answer."}
\vspace*{-.4em}
\end{quote}

\subsection{Brute-Force Behaviour}
There were a few cases (6/33, 5 \cond{without-narrative}, 1 \cond{with- narrative}) where participants visibly appeared to try and brute-force their way through some questions. P2 began brute-forcing solutions and ignoring feedback when exposed to histogram questions, but still read the story. For one histogram question, P9 read the question and associated choices but appeared to be unsure about the solution and tapped on all answers without reading the feedback box. P16 was observed to do this for several questions, while P18 did this for most questions. One participant (P13, \cond{with-narrative}) would occasionally make a random first selection just to see if the feedback box would give them a hint. \visdel{Regardless of whether or not participants engaged in brute-forcing strategies, a}\visadd{This said, a}ll participants appeared to read and think about each question. There were no instances observed where participants would immediately begin tapping on solutions.

When asked about whether \visdel{or not }they engaged in brute-forcing strategies, participants were \visdel{rather }honest in self-reporting \visdel{and would mention doing so }even if it happened on a single occasion. \visdel{We reviewed the recordings of the 7 participants who reported brute-forcing. W}\visadd{In our review of the sessions,} we saw that in many of these \visadd{self-reported }cases (5/7), participants read feedback and paused to think about what to pick next \visadd{ after}\visdel{, and that their brute-forcing was actually} some form of minor guessing at first -- in other words, children would report uncertainty and guessing as brute-forcing. The two participants (P16, P18) who were observed to engage in this behaviour admitted to brute-forcing during the interview. P18 did not elaborate on why they did so, but P16 stated:

\begin{quote}
\vspace*{-.4em}
\fontsize{8.5pt}{8.5pt}\selectfont
    \textit{``If I got it wrong the first time, I'd just press the other two. I just didn't think the other two were right when I got it wrong and I didn't want to go over [the question] again."}
\vspace*{-.4em}
\end{quote}

P16 also mentioned he did not notice the feedback box at all, \visdel{which }likely contributing to their brute-forcing behaviour. In observing how players interacted with the story, we saw that almost all participants took the time to read through the dialogue, with some exceptions \visadd{(P5, P16)}. %One participant (P16) occasionally switched between skipping and actively reading dialogue, but gave no indication as to why they did so. P5, on the other hand, clearly appeared to skip through the second half of the story in order to get to the educational activities faster and mentioned:
%\begin{quote}
%\vspace*{-.4em}
%\fontsize{8.5pt}{8.5pt}\selectfont
%    \textbf{P5:} \textit{``Here's the thing with the story and just giving questions right off the bat: you \emph{can} skip the story. You just have to read that question box and you have, pretty much, the question."}
%\vspace*{-.4em}
%\end{quote}

\subsubsection{Suggestions}
Participants in both conditions (6/16 \cond{without-narrative}, 3/17 \cond{with-narrative}) felt that the problems were unclear and lengthy at times, suggesting that we \visadd{should} shorten the question text. Many participants (15/33) said they would have liked to see more varied topics beyond \visdel{just }charts (e.g., fractions, percentages) or \visdel{a sort of level system where they could choose which }\visadd{the ability to target specific} grade level\visadd{s}\visdel{ to complete questions for}. \visdel{13 participants made suggestions related to}\visadd{On} story elements\visdel{:}\visadd{,} 7 \visdel{participants in the} (\cond{without-narrative}) \visdel{condition }wanted to see more of a story in the game, while 6 \visdel{participants in the} (\cond{with-narrative})\visdel{condition} would have liked a longer story line or branching narratives. Seven participants wished for more interactive activities, 5 would like to add character customization, and one thought it would be good to \visdel{have voices in the game}\visadd{include voiceovers}. 
 

%% file: 6-discussion-study.tex
\section{Discussion}
\label{sec:discussion}

\visadd{With the results of the study, we first address (RQ2): the impact of narratives on learning and engagement. With respect to learning, we detected a significant difference between the pre- and post-test scores for Q6 in the \cond{with-narrative} condition. Both Q5 and Q6 required participants to understand that the intervals on the x-axis should be equal. Given that a handful of participants mentioned they were not familiar with histograms, it could be that participants focused solely on the bars in the pre-test. After learning more about histograms and deception, participants may have looked at the graphs more carefully in the post-test. Beyond that,} while the overall impact on learning was not statistically significant, trends based on per-participant performance provide\visdel{d} \visadd{further} insight on the games and their effects on learning. 

\visdel{For example, the quantitative data showed that a}\visadd{A}ll children in the youngest age group either improved or exhibited no change in performance\visdel{ -- none of them performed worse on any questions in the post-test}, unlike in other age groups. This, combined with \visdel{qualitative feedback where children noted }\visadd{comments that} the game would be better suited for students \visdel{who are }new to a topic \visadd{and past work indicating that narrative text supports learning when prior knowledge of the content is low~\cite{Wolfe:2007}}, provides support for the game as a tool to support learning about new concepts. Moreover, it could be used to target children identified as struggling in a particular area or slightly younger audiences, though issues associated with reading comprehension skills may arise. \visadd{The imbalance in age/grade between conditions due to randomized group assignments could also mean less room for improvement in the \cond{with-narrative} condition.}%, swaying results in favour of the \cond{without-narrative} condition.} 

The usage of narrative elements appeared to negatively affect performance of male participants in our study, as well as exacerbate the degree to which people improve or do worse between the pre- and post-tests. The reason \visdel{for this }is unclear, but could perhaps be attributed to players becoming too focused on the story and not the educational content, \visdel{preferences stemming from gender differences}\visadd{gender-based preferences~\cite{Gros:2007, Carr:2009, Yee:2016, Yee:2017, ESAC:2018}}\visdel{ (in the case of the former)}, and/or fatigue. Participants in the \cond{with-narrative} condition spent nearly \emph{double} the amount of time playing their version of the game \visdel{because they had to read through dialogue and explore the game world}\visadd{due to the addition of dialogue and exploration views}, and we noticed one participant in this condition was visibly exhausted by the end of the study. This participant was also recorded to have performed considerably worse in the post-test. \visadd{As such, the results of the post-test and interview sessions in the \cond{with-narrative} condition may have been negatively impacted.}

On the impact of narrative elements on engagement, we saw significant differences in how participants rated their enjoyment of the game, specifically the art, activities, engagement, and fun. Higher ratings for the art were likely due to the presence of portraits \visdel{which}\visadd{that} appeared during conversations\visadd{, on top of the smaller sprites seen by those in the \cond{without-narrative} condition}. With respect to educational activities, some participants mentioned they appreciated how word problems were linked to the story and how the story helped make them feel invested in the problems they were solving. Meanwhile, improvements in engagement and fun could be attributed to the presence of exploration and story, giving players more to do than \visdel{simply }answering \visdel{math }questions\visadd{, supporting past findings on experiential play\cite{Wang:2009} and the benefits conferred through engaging with the material beyond simply reading and responding to questions\cite{Boeker:2013,Leutner:1993,Papert:1991}}. Overall, the inclusion of additional narrative elements had a positive effect on player enjoyment, while not impeding learning, which is encouraging. 

%% file: 7-design-space.tex
\section{Design Considerations for Educational RPGs}
\label{sec:design-considerations}

%\nath{this section does not really explicit the choice you have made. it sounds generic. I would structure it like this: 1) the choice YOU made. 2) why. so for each subsection you can structure it this way. eg: Target audience: We picked a small age range grade 1-3 because they can read and are learning similar concepts...}

Our game was motivated by design considerations grounded in game design and education literature. We intentionally limited the scope of our technology probe, so as to make the game amenable to an initial exploration of our research question \visadd{(RQ2)} in the context of a time-constrained study with child participants. While some guidelines for prospective game designers exist \cite{Salen:2003:RPG:1215723}, there is a lot of freedom afforded to designers with respect to what decisions they can make on various factors\visdel{, resulting in games taking on many different forms}. In this section, \visadd{we investigate (RQ1) and}\visdel{ we} present an overview of important factors for designers to consider when building an educational game for visualization literacy, the personal choices we made, as well as some alternatives and their associated trade-offs.

% Building on education and game design literature, we present a list of important factors to consider when designing an educational game focused on visualization literacy, and discuss associated trade-offs. %\fanny{cut -> }It should be noted that our list is by no means exhaustive -- such an endeavour is beyond the scope of our work.

\subsection{Target Audience}

Children of different ages are expected to have varying levels of reading comprehension, \visdel{which have large implications on the}\visadd{affecting the} narrative elements one can present. \visdel{Younger children are also more used to being read to as opposed to reading alone~\cite{Scholastic:2017}, which may influence a designer's choice in whether or not voice-overs are required. To test for reading level, we validated our content with professional educators, though another method could include analyzing the text using the Flesch-Kincaid readability tests~\cite{solnyshkina:2017}. Due to the heavy focus on reading in a narrative-based game, we designed our game for children aged 11-13 (i.e., grades 5-8) in an attempt to minimize impact on results due to differences in reading comprehension skills. We limited the age range so that learning tests and exercises were appropriate for all participants; we designed activities that would be most suitable to learners in grade 7 as we were interested in seeing whether younger (grades 5 and 6) and older (grade 8) learners may have a different response to the built prototype, as it is common for children to commence learning a novel concept in a grade, then revisit and further explore it in subsequent grades.} 
The target audience's prior familiarity of subject matter \visdel{is crucial in}\visadd{also plays a crucial role,} identifying what concepts can be built on and possibly consolidated, as well as which learning objectives should be addressed in the game. Children are exposed to visualizations as early as grade 1, starting with very concrete counting and sorting of physical objects, followed by pictographs, then highly abstracted visualizations such as bar charts~\cite{Alper:2017}\visdel{. When dealing with young children, however,}\visadd{, but it should be noted that} differences arise based on factors such as grade and location\visdel{, playing important roles in determining what one can present in an RPG-style educational game}.

% Children are slowly introduced to different types of visualizations over time, and the types of charts taught at each grade may vary between countries, districts, schools, and so forth. 
\visdel{For example, }A comparison of curricula shows that children in France~\cite{FranceCurriculum:2015}, Ontario~\cite{ManitobaCurriculum}, and various states \visdel{which}\visadd{that} have adopted the Common Core State Standards~\cite{USCommonCore} are expected to begin learning about data management in grade 1, whereas students in Manitoba~\cite{ManitobaCurriculum} begin in grade 2. With respect to \visdel{specific }chart types, the Ontario curriculum expects children to understand histograms and scatter plots by the end of grade 8, while the French curriculum mentions scatter plots (``graphiques cartésiens") and histograms as early as grades 3 and 4, respectively. %, whereas the Manitoba curriculum makes no mention of these types of graphs -- instead, they only mention circle, line, bar, double bar, and pictographs. 
Developing an understanding of how visualization \visdel{literacy }is taught at various ages and locations, so as to not make false assumptions about what children already know, is essential to inform the \visdel{type of }educational content to be covered.

\visadd{In our design process, we validated our content with professional educators. To measure readability, designers could also analyze text using the Flesch-Kincaid readability tests~\cite{solnyshkina:2017}. We designed activities that would be most suitable to learners in grade 7 to see whether younger (grades 5 and 6) and older (grade 8) learners may have a different response to the built prototype, as it is common for children to commence learning a novel concept in a grade, then revisit and further explore it in subsequent grades. From our study, we saw either improvements or no change in younger grades, but this was not true for older groups. This does not mean, however, that we should solely include content from higher grades --- some children are bound to struggle with concepts more than others. Personalized experiences (e.g., adjustable difficulty, topic/concept selection) would be helpful here, though more work is required to further explore this avenue.}

\visdel{In grade 7, pie charts should already be quite familiar to learners, whereas histograms would be a newer concept -- we were interested in exploring both in our game to evaluate the probe as both a tool for teaching new concepts and refining already-acquired skills.} %\fanny{Add a sentence about variety of problems covered; as well as activities perhaps more challenging (given that Grade 7 activities were found "too easy" by more experienced children?}

% younger children (11-year-old) => improved / no change
% some (the few who seemed really confident in histograms) wanted to see more challenging content,

%\fanny{so? what did you pick? how?}

\visdel{Children of different ages are expected to have varying levels of reading comprehension, which have large implications on the narrative elements one can present. We validated our content with professional educators, and another method could include % whether or not the language they use is appropriate for the targeted grade e.g., by 
analyzing the text using the Flesch-Kincaid readability tests~\cite{solnyshkina:2017}. Younger children are also more used to being read to as opposed to reading alone~\cite{Scholastic:2017}, which may influence a designer's choice in whether or not voice-overs are required.}

%Moreover, children of different age ranges have varying propensities for whether they would choose to read for fun in their spare time, and how familiar they are with electronic games\cite{Scholastic:2017}. The former may affect how engaged children might be with narrative-focused games, while the latter could affect how easily they can pick up the game mechanics, and thus the need to create tutorials to teach how to play.% the game.

\subsection{Time}
\visadd{On top of the time spent on the educational content, designers must also consider how much time is needed for players to interact with the narrative elements. In our study, there was not a significant difference between the time spent on educational activities between the two conditions, but the \textit{total} time spent in the \cond{with-narrative} condition brought the average play time to double that of the \cond{without-narrative} condition. This disparity cannot be understated, but the boost to engagement without sacrificing learning shows that the inclusion of narratives can be beneficial. Still, designers must be considerate of the trade-off between time and engagement, and whether it is ``worth it'' in the context of their own games. Creative game mechanics where the narrative is presented mostly in the form of expressive graphics (e.g., as found in the game Florence~\cite{florence}) are worth investigating to mitigate time cost.}

\subsection{Educational Activities}
In school, children are given various types of problems involving data visualizations~\cite{Alper:2017, shreiner:2018} (see Table~\ref{tab:questions}), \visdel{each of which} target\visadd{ing} different cognitive resources and requir\visadd{ing} development of a complementary set of skills, from thinking critically about different options to choose from, to generating explanations for one's reasoning. Ideally, an educational game focused on visualizations should leverage all types of problems borrowed from current best practices and research in education, in order to support holistic competence development. For instance, prior research has shown that response format (e.g., multiple choice) is not nearly as important as stimulus format (i.e., context-free or context-rich) \cite{Schuwirth}. 

We opted for multiple-choice questions due to the ease of implementation, verification, and their ability to support a breadth of topics, and looked at sample exercises and worksheets \cite{TIPS4RM, IXL, KhanAcademy} for guidance. We noticed many chart questions could be classified further into either create a chart-type questions, or interpret the chart-type questions. The latter would be simple to translate into multiple choice questions, but the former would present a challenge unless players are asked to make a choice for each part of the chart (e.g., ``Which of the following options is best suited as a label for the y-axis in this chart?"). To address this, we made use of questions where players would be presented with different charts and have to choose the one \visdel{which}\visadd{that} best fits a scenario, followed by a question requiring to interpret the graph, allowing learners to bring resolution to the problem at hand. 

%\elaine{this is the first sentence of the previous paragraph, isn't it?}
%For implementation and evaluation simplicity, we limited our activities to include context-rich, multiple-choice questions that could be easily validated computationally. 

%\fanny{so, what did you end up picking? this is why we chose a variety of input ...}

\visdel{Our technology probe was limited in its scope by design. However, }Solving problems related to data visualizations encompasses a variety of activities, ranging from data collection, data cleaning and wrangling, data manipulations, and choice of mappings to general considerations about how to best represent the same data to address specific questions, think critically about the generated visualization, and ultimately communicate findings. Our game started by tackling data-related problems directly translated and adapted from textbooks. Developing a game \visdel{which}\visadd{that} encompasses all of these activities goes beyond the scope of our paper, but we share our reflections on how the inclusion of different activities could guide the design of a narrative-based educational game more generally \visadd{(\S\ref{sec:lessons})}.

%From our review of previous survey research and various lesson outlines, worksheets, and exercises \cite{OntarioCurriculum:2005, IXL, KhanAcademy} we distill an initial set of question types seen in data-related problems (Table~\ref{tab:questions}).
% However, given the presence of an overarching narrative, all questions in an educational RPG could easily be context-rich if designers relate the questions directly to the story. In this section, we discuss the trade-offs of common question types to help inform designers, and discuss each with respect to facets such as validation and feedback, brute-forcing, and time to complete activities.

\begin{table}
\small
\begin{tabular}{r p{0.7\linewidth}}
\textbf{Chart} &  Students are asked to create a chart \visdel{(typically from scratch) }to represent a given set of data \visdel{. Charting questions generally require students to}\visadd{and oftentimes} also label parts of the chart. \\ \hline
\textbf{Label} &  Students are given a chart or table and asked to add labels (e.g., column names, axes, categories, intervals). \\ \hline
\textbf{Match} & Students are provided with sets of entities (e.g.\visadd{,} \visdel{a set of words and a set of descriptions, or} a set of charts and a set of sentences/inferences) and asked to and asked to determine a 1:1 mapping of entities from one set to another.\\ \hline
\textbf{Multiple Choice} & Students are given a question and a set of choices, and are required to identify the most-correct response to the question. \\ \hline
\textbf{Numeric} & Students are given a question requiring them to read a chart or table and report a single number to answer the question\visdel{ (e.g.\visadd{,} the total number of respondents in a particular category)}. \\ \hline
\textbf{Plot} & Students are given a chart (either incomplete or an empty template) and asked to plot data points or populate it with data. \\ \hline
\textbf{Survey} & Students are tasked with surveying others and collecting data.\\ \hline
\textbf{Discussion} & Students are given a topic and asked to verbalize their thoughts in a classroom setting, hearing and responding to input from fellow students and facilitators (e.g., teachers).\\ \hline
\textbf{Short Answer} & Students are asked a question about some provided data or visualization and required to give a short textual response.\\ \hline
\textbf{Long Answer} & Similar to short answer questions, but more complete sentences \visadd{(and possibly justification)} are expected . \visdel{Long answer-type questions may or may not require students to provide justification for their answers.}
\end{tabular}
\vspace*{0.3em}
\caption{Types of questions for data-related problems.}
\vspace*{-3em}
\label{tab:questions}
\end{table}
%

%\subsubsection{Validation and Feedback}
One of the challenges with stand-alone educational games lies in validating answers and providing appropriate guidance and feedback to the student to support learning. The variety afforded by having humans validate responses in traditional settings does not translate well to games -- we are restricted in how we can automatically assess answers, affecting the types of questions we can administer to players \visadd{(Table~\ref{tab:questions}}), as well as the level of feedback. %Depending on the complexity of the data problem posed and targeted learning objectives, different formats can be used\visdel{: closed (plot, match, multiple choice, numeric), open-ended (discussion, short answer, long answer), and mixed (label, survey, chart)} \visadd{(Table~\ref{tab:questions}}).
Responses to closed questions are very easy for computers to validate, as solutions are largely binary\visdel{ -- they are either correct or incorrect}. Such questions are suitable to visualization tasks where a clear, unique answer is possible (e.g., retrieve a value, basic comparisons of two elements, identify an outlier)~\cite{amar2005low}. While the rigidity of these question formats facilitates implementation and assessment, it also acts as a catalyst for the shortcomings. Numeric- and plot-type questions are limited in what skills they can test and are primarily used for testing basic chart- or table-reading\visdel{, respectively}. \visdel{Meanwhile, t}\visadd{Also, }the answers for multiple choice and match-type questions are always visible to learners, making them ill-suited for topics \visdel{which}\visadd{that} may require spontaneous generation of responses \cite{Schuwirth}.

Open-ended questions afford more nuanced, richer forms of input \visdel{which}\visadd{that} are often necessary for the educators to understand the learner's reasoning process\visdel{ behind their solution}. For data-related problems with visualizations, it is possible that different analytical paths yield equally reasonable answers, especially when patterns are subtle, data is uncertain, or the problem can be interpreted differently (i.e., in a problem asking ``what makes these shops successful?", the concept of success could be ambiguous or hold multiple meanings%could possibly be operationalized differently
). Thinking critically about how to approach a problem and use data and visual representations of this data has been highlighted as an essential component of visualization literacy~\cite{Chevalier:2018}. Open-ended questions, though typically requiring sophisticated methods of validation, such as %through the employment of 
natural language processing techniques (e.g., ~\cite{hirschman:2001}) or computer vision methods (e.g., color mapping extraction~\cite{poco:2017} to be used by intelligent tutor agents (e.g., ~\cite{rivers:2017})) should be considered in more advanced, comprehensive educational games. 
%One automatic method of validation could be to do a search for a set of keywords, but this could result in issues if learners are familiar with alternative terminology, or cannot recall the exact words. 
Further effort is required to create a platform that (1) supports interactions between students and teachers, (2) provides teachers with a way to assess the quality of discussions, and (3) allows for seamless transitions back to the rest of the game, so as to not impede on player immersion.

%Mixed-type question can be reclassified as closed- or open-ended, depending on how one envisions implementation. %Validation of label questions is simple if learners are provided with a word bank or drag-able label objects. Without such features, label questions are similar to a series of short answer (complex) questions. 
%Survey questions can be made simple if integrated into the game world (e.g., requiring players to talk with characters to collect data). If real data is required, a human facilitator is likely needed to assess the quality of the data, making validation complex. Chart-type questions are %, in their natural form, 

%difficult to automatically validate, and would require. However, if decomposed into a series of plot and label (simple-variant) problems, they can be handled simply.

%\subsubsection{Other considerations}
Other facets to consider include time needed to complete activities, as well as how prone different formats are to brute-forcing strategies. For instance, multiple choice questions can be very quick to answer, but are considerably more vulnerable since learners are given immediate feedback and may exhaust all possible solutions until they pass. A good balance must be struck between time spent on activities compared to the rest of the game, as well as the variety of problems. If players are constantly required to answer open-ended questions every few minutes, they may quickly grow bored or frustrated by the game. \visadd{Designers could also explore other input methods to see if they add to playfulness (e.g., Florence~\cite{florence} uses shake, drag, and draw instead of simple clicks)}.

\subsection{Narrative Design}
\label{sec:narrative-design}
Data-related problems are commonly presented with stories to provide context and motivation for resolution of the problem. To the best of our knowledge, \visadd{while other educational games with narratives exist\cite{Prodigy, flocabulary, brainpop}, }this work is the first to leverage narratives to inform the design of an \visadd{RPG-like} educational game focused on visualization literacy, and evaluate whether such an approach would be conducive to learning.  %\nath{stronger here - this is main novelty for vis i think: We are the first one to leverage narrative to blablabla and to evaluate whether this will be effective.} 
%\fanny{Need a couple sentences talking about our narrative-ish}
%As such, we see an opportunity to leverage narrative elements in an educational game focused on visualization literacy. \elaine{isn't this more of an intro thing? it gave us a way/point to justify exploring stories in educational games for vis. lit.} 
With the exception of larger projects \visdel{which}\visadd{that} span over several days and cover a range of activities, individual exercises in textbooks and worksheets are typically presented as being independent and disconnected from one another, making it difficult for a design to translate and piece these elements into a singular cohesive and engaging narrative. Below, we address a few topics related to narrative design.
%\nath{this next paragraph is redundant, i read it before about the whole range of activities. cut. plus its good to focus on the novel contribution here. You are the first one gamifying VIS and incorporating narrative elements in the game. I would create a short paragraph that explains in bullet point the key narrative components you have introduced that you are going to talk next: model/character etc}
%While our example game is limited in its scope to enable initial exploration of our research questions, future games should strive to bear inspiration from the way longer course projects are designed to address a larger problem in piecemeal. In our context, this could include initial activities consisting of collecting data first, through exploring the game world and interactions with its inhabitants; the player could then be required to attempt at solving the problem, and discover some challenges, e.g. data need to be cleaned, requiring to go on quest to acquire the relevant compentency for say, removing erroneous data and outliers. A compelling story can be weaved around visualization objectives, offering a variety of educational activities motivated by characters having to solve problems in the game, while providing a playful experience. Below we address a few topics related to narrative design.

\subsubsection{Narrative Model}
There are \visdel{multiple}\visadd{many} ways of \visdel{combining together a set of}\visadd{integrating} activities \visdel{within}\visadd{into} a narrative. In games, narratives often follow one of three main models: linear, string of pearls, or fully-branching \cite{Gamasutra:2019}. Linear models are very rigid, requiring players to follow a specific sequence of events, which represent an appropriate choice if the activities encompassed in the game must all be completed by the learner to acquire the targeted competencies. Because different learners may require different emphasis on certain topics, or some more advanced activities may be provided for a motivated learner to learn beyond the core set of exercises supported by the main story line, additional quests could be integrated, following what the game design community refer to as string of pearls. In RPGs, this is often accomplished through side stories or quests that allow players to take a break from the main story line to interact with other characters and learn more about the game world. Fully-branching games push this even further, allowing players to make choices that directly impact the story and results in players being shown different events given previous decisions. The latter is particularly valuable in instances where a learner requires more practice on a particular topic, as evidenced by the number of attempts or amount of guidance required to solve the core exercises. 

Our design probe employed a linear model, as we wanted to force each player to go through the same set of steps, making for fairer comparisons between subjects in an empirical study, as opposed to other models that afford more flexibility, and therefore, possible confounds. Future work should explore how different narrative models support different needs and preferences. 
%\nath{again, what model did you choose here? how is it tailored to kids? and to Vis lit?}
On a related note, it is worth mentioning the amount of freedom and agency offered by the different models may also impact the enjoyment and player experience, as making players feel as if their choices matter acts as a psychological motivator \cite{Lee:2006, Ke:2013}. % The choice of model also bears implications on the resources required (e.g., writing, art), as well as how the game should be evaluated from a research standpoint, with linear narratives 

%On the other hand, comparisons between subjects exposed to games with string of pearls and fully-branching models can be complicated, as it would be difficult to ensure all subjects are exposed to the same amount of learning material for testing purposes. There are also many confounds to consider when using these models, such as time spent on exploration, quests, or character interactions. For example, if subjects are given 30 minutes to interact with a game, one may spend the majority of their time completing quests related to bar charts, while another may focus solely on advancing through the main story line, possibly resulting in the latter acquiring more breadth but less depth of chart topics.

\subsubsection{Characters}
%\nath{shorten; i want to know what you picked. maybe explain that you wanted to limited the complexity of the game to enable kids to focus?}
\visdel{Another point of consideration for narrative design is the presence of characters. }In word problems, questions are usually presented in a way that introduces a character and the problem they are facing. With an overarching narrative, an educational RPG can spend some time presenting background information for characters, giving further motivation as to why their problems are worth solving. \visdel{We created a set of nine different characters, including the main character that the player embodies. }In creating characters, designers must be aware of how character design and presence affects factors such as impact on player immersion, attachment or identification~\cite{cohen:2013, konijn:2009}, cognitive and affective learning~\cite{paiva:2005}, resources required, and time spent on exploration and interactions. We strove for diversity of gender and personality, but opted for simple character stories in order to minimize time required to meet each character. \visadd{Participants noted the lack of depth of characters, and that some were only met once. Designers could consider using characters from established universes, as characters familiar to the learner could reduce introductory text, and would build on an already developed attachment and empathy for these characters.}

\subsubsection{Dramatic Arc}
Freytag's dramatic arc model \cite{Freytag:1863} has been described as ``the heart of any good drama and any good game system" \cite{Wang:2009}. This arc consists of five parts: the exposition/introduction, rising action, climax, falling action, and catastrophe/denouement). Many stories tend to incorporate this arc, but vary in how much time they spend on each phase, as well as the number of times they repeat a phase or set of phases. Prior studies in education and psychology have shown that stories \visdel{which}\visadd{that} follow a dramatic arc are more likely to elicit empathy for characters in adult audiences \cite{Barraza:2015}, and are more well-received by children when compared to stories without dramatic arcs \cite{Guillot:2013}. It should be noted that while including a dramatic arc may increase engagement with the narrative, its effects on learning are unclear and remain to be explored. 

%\nath{i dont see why we need this next paragraph. cut}
%In games where the narrative is not the main focus, designers may choose to overtly provide some backstory through means such as game manuals or introduction sequences \cite{Juul:2001}, or hint at a theme through the design of assets (e.g., cutesy cartoon monsters %and flashy magic spells 
%would be received differently from one with %dark visuals and 
%vs. highly-realistic character models.) 
%In contrast, RPGs -- where the gameplay is largely plot-driven -- tend to require more emphasis placed on crafting interesting stories and supporting narrative elements. In this section, we address a few topics related to narrative design.

\subsubsection{Genre}
Whether or not a story is interesting to an individual is largely due to personal preferences and very difficult to account for -- it is \emph{not} possible to create a story that appeals to all audiences. A survey looking at player motivations found that gender disparities exist when looking at genres, e.g., females were found to be more interested in games with high fantasy themes as opposed to science fiction \cite{Yee:2017}. Such findings may be used to help guide the creation of narratives.
%but designers must be aware that these are generalizations and they cannot realistically be expected to appeal to all members of a particular demographic.

There are many different genres of stories (e.g., science fiction, high fantasy, mystery). For brevity, we generalize and discuss two groups: fantasy and non-fantasy. Fantasy is a large part of the lives of newer generations \cite{Prensky}, and is considered an intrinsic motivator \cite{Lepper:1987}. A study where children were offered the chance to interact with a game in either a fantasy or no-fantasy context, found that children preferred to choose fantasy \cite{Parker:1992}. Moreover, RPGs generally make use of fantasy stories, so if the goal is to make an educational game similar to commercial ones (e.g., perhaps to give the illusion that it is more of a game than a learning tool), it may be intuitive to make use of fantasy elements. However, there are concerns regarding its impact on education.

Some literature in education suggests that fantasy hinders the ability for young children to transfer knowledge from fictional settings to real-world scenarios and, instead, recommends realistic content \cite{Richert:2009, Richert:2011, Ganea:2014, Walker:2015, Nyhout:2018}. Designers must consider the trade-offs between education and motivation when choosing a genre. Furthermore, past work has cautioned care in ensuring that usage of fantasy elements does not compete with educational content for player attention, as increased motivation with the context may negatively affect learning \cite{Lepper:1987}.

\subsection{Game Design}
\visdel{Our design probe adheres to a simple RPG design supporting exploration and character interactions. }We \visadd{now} discuss points of consideration in the design of the game \visadd{specifically}. In weighing options\visdel{ for each section}, care must be taken to ensure that the game adequately supports and facilitates the integration and interleaving of educational elements with narrative elements, in order to address learning objectives whilst not interrupting a player's immersion \cite{Veloso:2013, Westera:2019}. \visdel{In our probe, battles were replaced with educational activities, and voices, multiplayer aspects, and other motivators (e.g., currency/reward systems and customization) were omitted because we did not consider them pertinent in determining whether or not narrative elements affect learning and engagement.}

%\subsubsection{Target Platform}
%Popular platforms for games include desktop computers or laptops, tablets, smartphones, and consoles (either home systems such as PlayStations or handhelds such as the Nintendo 3DS). Consoles typically require a more extensive approval process before games are made available on their systems -- as such, designing a game targeting these platforms for a short study may not be viable, nor worthwhile. In looking at other platforms, mobile gaming has become increasingly popular across children in North America \cite{NPD:2013, NPD:2015, ESAC:2018}, adding to the appeal of designing for tablets and smartphones. When comparing these platforms, past work has shown that the usage of tablets outperforms personal computers when used as supplementary learning tools\cite{Papadakis:2018}. Furthermore, when compared to smartphones, children are more likely to use tablets on a daily basis due to the ease of use afforded by increased screen sizes \cite{Lillard:2017}.
% \fanny{Add other countries? or even find data worldwide, or western populations?}
% \elaine{Stats for mobile gaming in Europe (2015), but not specifically about children: \url{https://www.isfe.eu/wp-content/uploads/2018/11/deloitte_report_isfe_2015.pdf}}

\subsubsection{Feedback}
Regarding the educational aspect of the game, designers must consider how they want to convey to players whether they have done something correctly or incorrectly, as well as how much information or feedback to present when a player answers a question. Prior research has shown that support features such as advice is not conducive to learning, whereas feedback and just-in-time information promotes learning \cite{Leemkuil:2005}. Designers must consider, then, how much feedback is to be presented, whether or not the feedback is meaningful, as well as how to display this feedback. Preset feedback messages\visdel{ (e.g., a message to be displayed when a player answers a question correctly or not)} are simple to implement, but require designers to anticipate reasons for why a player might have answered a question incorrectly. \visdel{A more intelligent system may look at previous actions taken by the player and craft more personalized feedback, but would certainly require more of an effort to implement.}

\visadd{In our study, children were mostly positive about our method of presenting feedback, but some noted they would occasionally skim or ignore feedback because they believed it would not tell them anything they did not already know. Designers should acknowledge that children may ignore the feedback attached to correct answers, or that they will only selectively read up to the point where they realize what they did wrong. To improve reading rates or engagement with feedback in general, designers could explore the efficacy of personalized feedback, though this would require more effort to implement.} %. A more intelligent system may look at previous actions taken by the player and craft more personalized feedback, but would certainly require more of an effort to implement.

\subsubsection{Exploration}
Many RPGs tend to have some sort of exploration involved, allowing players to move around the fictional world at their own pace. \visadd{Exploration is not commonly leveraged in educational games including narratives~\cite{Prodigy, flocabulary, brainpop}}. The ability to freely roam and explore creates the illusion of the game being less linear than its narrative-driven gameplay may imply \cite{Carr:2003}. Thus, exploration should not be overlooked in the game design. Designers should first consider how movement is handled in the game: do players simply tap on hotspots to load a connected map and start dialogue (like in point-and-click adventures), or do they see and control characters moving around the actual game space (like in\visdel{more} modern-day RPGs)? The former saves time dedicated to \visdel{animating character movement and specifying obstacles and colliders for models/sprites}\visadd{animations and defining colliders}, but may not feel as immersive to players since they are not able to explore as freely\visdel{ as in the latter option}. One must also consider the size of the world and each area or map: a more expansive set of maps may make the world seem more open and free, but may run the risk of players becoming lost, on top of requiring more time for travel.

\subsubsection{Combat}
Combat or battles are often used in RPGs as a way of challenging and testing players\visadd{, as they must improve and develop tested skills to progress.} Players are encouraged to complete side quests and defeat lower-leveled enemies in order to accumulate more experience and items to become stronger and face harder encounters. In an education setting, this system could be used in a few ways: players could complete educational activities to unlock skills or items to make battles easier, players could be asked to complete educational activities in order to inflict damage to enemies in combat, \visdel{or the }battles could \visadd{be where players collect items~\cite{Prodigy}, or} be replaced by educational activities entirely. \visdel{In all cases, players are required to improve and develop the tested skills in order to overcome challenges and progress through the game.}

\subsubsection{Voices}
Voice acting is commonly used in commercial games to further take advantage of auditory streams and add depth to characters\visdel{ (if they exist)} and/or the narrative. In an educational game, voices can be used to help children follow along with the text. \visdel{As previously mentioned, }Younger children are used to having someone read to them \cite{Scholastic:2017} and the presence of voices could help emulate this environment\visdel{. However}, \visadd{but} voice acting requires\visdel{the acquisition and implementation of} additional assets, which may be costly depending on the \visdel{desired }level of professional quality. %Designers must also consider how much of the game should be voiced: is it only character dialogue that gets voiced, or are all questions also read aloud to players?

\subsubsection{Multiplayer Aspects}
Some games support multiplayer capabilities through methods such as competition (e.g., player vs. player, leaderboards), socialization\visdel{ (e.g., chat features)}, and/or co-operative play. Findings regarding competition in educational games appear to be mixed. Some found that competitive elements increased learning and motivation \cite{Cagiltay:2015}, while others found the opposite\visdel{ -- participants in a no-competition group performed better on post-tests than those in the with-competition group}~\cite{Chen:2018}. 

%\nath{maybe you can cite my paper on peer learning :) https://dl.acm.org/doi/abs/10.1145/2998181.2998231}
Socialization or discussion with others has been shown to support learning \cite{Leemkuil:2005, sobel2017edufeed}, but when dealing with young children, designers must consider ways of moderating discussions, either automatically or by requiring a an educator or other adult to monitor the chat to ensure the space remains safe for students and conducive to learning. Co-operative play could be used to support collaborative learning, but a game designed specifically for groups could run into issues where players are unable to find others to play with, or simply do not enjoy connecting with others for the sake of an educational game. %More effort also needs to be expended to consider how to modify all mechanics and aspects in the game to make use of the additional number of players. 

For all of these methods, designers must also consider\visdel{the added technical requirements for properly supporting multiplayer features, as well as} whether or not supporting these features in the game is actually needed. In a classroom setting, teachers could emulate competition, communication, and co-operative play by setting up in-class leaderboards, holding classroom discussions, and allowing students to play together. Finally, % one must also understand that 
the impact of these features on individuals can be varied due to personal preferences. For example, the presence of competitive elements may motivate those who are naturally competitive, but could add unnecessary pressure and drive away those who are not. 

\subsubsection{Other Motivators}
Games make use of various motivators \visadd{(e.g., customization, achievements) that} \visdel{to captivate their audience and there are many features}we did not address\visdel{, such as customization and achievements}. Designers may be interested in adding these features into an educational RPG to enrich it, but the effects of each motivator on learning has largely yet to be explored. %As previously mentioned,
As personal preferences play a large role in how effective each motivator is for an individual\visadd{\cite{Yee:2016}}, it can be difficult to anticipate the effectiveness of including certain motivators. \visdel{For instance, some players may be more drawn to free exploration, whereas others are enticed by fast-paced action and battles\cite{Yee:2016}.} \visadd{Unsurprisingly, our participants expressed an array of opinions on all aspects of the game due to personal preferences: some suggested adding more competitive elements, while others noted disdain for such features. On the educational aspect, there were conflicting answers on how much to focus on education versus play, as well as topics to concentrate on or cover. Such comments speak to a need for more personalized experiences and/or additional features that children are not required to interact with (e.g., optional puzzles).}

%% file: 8-lessons-learned.tex
\section{Lessons Learned: Considerations for Design}
\label{sec:lessons}

%\visadd{We reflect on lessons learned and highlight gaps to be addressed in future works in the area of gamification to support visualization literacy.}

%\visadd{\subsection{On Games for Information Visualization}}
%\visadd{\textbf{Visualization affects design.} The design of our probe was driven largely by the learning objectives we covered (i.e., chart reading and interpretation). A multiple choice setup allowed us to quickly address and test a variety of topics, but consider the case where a designer may choose to focus on chart synthesis instead -- a designer may seek inspiration from mechanics found in sandbox games, wherein players freely build and destroy structures. Similarly, a game focused on exploration of data may draw from features seen in exploration games. From this, we note the impact of learning objectives on game design.}
\visadd{\textbf{Visualization learning objectives affect game design.}
The design of our probe was driven largely by the learning objectives we covered (i.e., chart reading and interpretation). A multiple choice setup allowed us to quickly and robustly address and test a variety of topics, but consider cases where designers may instead choose to focus on visualisation construction~\cite{Bishop:2020} or exploration of data; the gameplay would have to accommodate such activities seamlessly, seeking inspiration from mechanics found in existing games where possible (e.g., in Metrico~\cite{metrico}, player actions such as jumping and walking impact the visualization). We brainstormed several possible gameplay options before fleshing out the learning objectives, and found that such an approach constrains activity design, potentially limiting their educational value -- in other words, forcing activities to comply to game mechanics, rather than prioritizing the quality of the educational content, could hinder the learning experience. Instead, we recommend that designers first focus on activities \textit{independent} of game mechanics and visual design, as we believe one of the main dangers in designing educational games is to focus more on play rather than learning.}

\visadd{In gamification, this balance between learning and play is often explored in \emph{stealth learning}~\cite{sharp2012stealth}. While this was not our focus, we do briefly touch on ways that activity and game designs could accommodate stealth learning through immersion and other means (\S\ref{sec:design-considerations}). Future work performing a systematic analysis of visualization learning objectives specifically, along with educational activity types and suitable game mechanics would both aid future designers, as well as help organise research efforts investigating what strategies best support learning of different visualization literacy skill sets.}

%\visadd{We believe one of the main dangers in designing educational games is to focus too much on supporting \emph{stealth learning}~\cite{sharp2012stealth} through gamification -- that is designing activities to fit within a game for learning in disguise, rather than focusing on relevant activities that may feel like studying but better support learning. We briefly touch on activity and game design (\S\ref{sec:design-considerations}), but future work performing a systematic analysis of visualization learning objectives specifically, along with educational activity types and suitable game mechanics would both aid future designers, as well as help organise research efforts investigating what strategies best support learning of different visualization literacy skillset.}

\visadd{\textbf{Game design affects visualization design.}
The relationship between visualization and game design is, by no means, unidirectional. The design of the game also shapes the visualizations. Designers may see a need to move \textit{away} from best practices previously established by the visualization community in order to better assimilate charts and other such assets within their game. For example, the usage of sharp graphics (as opposed to sketchy graphics~\cite{liu2014survey}) or certain colour schemes may clash with the aesthetic of the other art assets. The clear distinction between ``fun'' assets and ``serious'' assets could break a player's immersion in the game environment and impact their learning and engagement. In our game, we opted for sketchy rendering for the 
choose-a-chart questions intentionally as the presented options were to be seen as visualization templates, rather than representative of the underlying data. This choice influenced the design of the interpret the graph questions, which we decided to make consistent, though this may impact reading and interpreting the graphs. Future work should explore such trade-offs in the context of information visualization specifically.}

%% file: 9-limitations.tex
\section{Limitations}
Games are usually played at one's own pace and players are usually able to stop and resume playing at different periods throughout the day or across multiple days. In our study, some participants reported playing games for less than one hour per week, but we required participants to focus solely on the game for upwards of 30min in a single session, leading \visdel{in}\visadd{to} fatigue. A more true-to-life evaluation would perhaps allow players to play in shorter bursts across a longer period of time. This would also address cases where participants felt characters were shallow or that a longer story would be better, as it would give us more freedom to extend the narrative to create a more rich and immersive experience. 

The design of our pre- and post-tests were also limited \visadd{despite being co-designed with educators}\visdel{despite having sought feedback from an educator prior to conducting tests}. We opted for multiple-choice questions to save time and have a test well-aligned with our games, but it is unclear whether changes in performance between the two tests were due to actual changes in knowledge or the result of guessing. Open-ended questions may have given us more insight into how much information they acquired from the game, but the \visdel{trade-off on }time needed to answer questions would be problematic for a one-hour study. The limited sample size was also \visdel{problematic}\visadd{an issue}; studies were conducted during the active school term when students and guardians have busy schedules, and concerns surrounding COVID-19 resulted in early termination \visadd{of study sessions}\visdel{of recruitment and cancellation of sessions}.

There are also limitations caused by the design of our game. \visdel{As previously mentioned, there are many factors to consider when designing a game and its narrative. }Our narrative used a fantasy world, but this may only appeal to some children. \visdel{Others may be more interested in science fiction-type stories, or perhaps topics situated in reality. So, t}The implications of our narrative elements on learning and engagement may not necessarily apply to other genres; \visdel{and} more could be done to compare non-fantastical \visdel{worlds }to fantastical worlds, \visdel{as well as}\visadd{and} variations of each and how suitable they may be to address \visdel{different }visualization learning objectives. %\fanny{what can we say here about graph types or visualization activities?}

%% file: 9-conclusion.tex
\section{Conclusions}
\label{sec:conclusion}
In our work, we looked at how gamification could be used to support visualization literacy education and contributed what we believe to be the first story-based game \visdel{to support visualization literacy}\visadd{in this space}. Results of our evaluation with child participants were encouraging with regards to engagement, though results on learning were mostly inconclusive. Qualitative feedback showed many conflicting opinions on various topics (e.g., characters, \visdel{inclusion of features such as}combat), \visdel{the disparity of which echoes the}\visadd{echoing} differences in past findings regarding narratives and learning performance \cite{Adams:2012, Bittick:2011}. \visdel{We posit that this variety is largely due to the overwhelming number of factors involved in narrative and game design. }\visadd{We outlined considerations for narrative and game design, and}\visdel{We} see value in further research studying the impact of \visdel{different aspects of narrative and game design}\visadd{each aspect} (e.g., narrative genres, motivators), \visdel{and}\visadd{as well as} how these can be best used \visdel{at the service of}\visadd{to augment} visualization teaching and learning, and hope our work will inspire future explorations in this promising area. 